\documentclass[twocolumn,english,showpacs,preprintnumbers,amsmath,amssymb,floatfix,nofootinbib]{revtex4-1}

\usepackage[T1]{fontenc}
\usepackage[latin9]{inputenc}
\usepackage{color}
\usepackage{array}
\usepackage{amstext}
\usepackage{graphicx}
\usepackage{esint}
\usepackage{rotating}
\usepackage{slashed}
\usepackage{siunitx}
\usepackage[normalem]{ulem}
\usepackage[font=small,labelfont=bf]{caption}
\usepackage{xcolor}
\usepackage[colorlinks = true,
linkcolor = magenta,
urlcolor  = blue,
citecolor = red,
anchorcolor = blue]{hyperref}

\newcommand{\pom} {I\!\!P}
\newcommand{\reg} {I\!\!R}
\newcommand{\xpom}{x_{\xpom}}

\makeatletter

\providecommand{\tabularnewline}{\\}

\@ifundefined{textcolor}{}
{%
 \definecolor{BLACK}{gray}{0}
 \definecolor{WHITE}{gray}{1}
 \definecolor{RED}{rgb}{1,0,0}
 \definecolor{GREEN}{rgb}{0,1,0}
 \definecolor{BLUE}{rgb}{0,0,1}
 \definecolor{CYAN}{cmyk}{1,0,0,0}
\definecolor{MAGENTA}{cmyk}{0,1,0,0}
 \definecolor{YELLOW}{cmyk}{0,0,1,0}
 }

\@ifundefined{definecolor}
{\usepackage{color}}{}
\@ifundefined{definecolor}
{\usepackage{color}}{}
\makeatother
\usepackage{babel}

\def\Re{{\cal R \mskip-4mu \lower.1ex \hbox{\it e}\,}}
\def\Im{{\cal I \mskip-5mu \lower.1ex \hbox{\it m}\,}}

\def\tev{\,{\ifmmode\mathrm {TeV}\else TeV\fi}}
\def\gev{\,{\ifmmode\mathrm {GeV}\else GeV\fi}}
\def\mev{\,{\ifmmode\mathrm {MeV}\else MeV\fi}}
\def\to{\rightarrow}

\begin{document}

%
%
\title{A global QCD analysis of diffractive parton distribution function considering higher twist corrections within the xFitter framework} 
%
%

\author{Maral Salajegheh$^{1}$}
\email{maral@hiskp.uni-bonn.de }

\author{Hamzeh Khanpour$^{2,3,4}$}
\email{Hamzeh.Khanpour@cern.ch}

\author{Ulf-G Mei{\ss}ner$^{1,5,6}$}
\email{meissner@hiskp.uni-bonn.de }

\author{Hadi Hashamipour$^{4}$}
\email{ H\_Hashamipour@ipm.ir}

\author{Maryam Soleymaninia$^{4}$}
\email{Maryam\_Soleymaninia@ipm.ir}

\affiliation {
$^{1}$Helmholtz-Institut f\"ur Strahlen-und Kernphysik and Bethe Center for Theoretical Physics, Universit\"at Bonn, D-53115 Bonn, Germany  \\
$^{2}$Dipartimento Politecnico di Ingegneria ed Architettura, University of Udine, Via della Scienze 206, 33100 Udine, Italy \\
$^{3}$Department of Physics, University of Science and Technology of Mazandaran, P.O.Box 48518-78195, Behshahr, Iran  \\
$^{4}$School of Particles and Accelerators, Institute for Research in Fundamental Sciences (IPM), P.O.Box 19395-5531, Tehran, Iran   \\
$^{5}$Institute for Advanced Simulation and Institut f\"ur Kernphysik, Forschungszentrum J\"ulich, D-52425 J\"ulich, Germany \\
$^{6}$Tbilisi State University, 0186 Tbilisi, Georgia
}

\date{\today}

%
\begin{abstract}

We present {\tt SKMHS22}, a new set of 
diffractive parton distribution functions (PDFs) and their 
uncertainties at next-to-leading-order accuracy 
in perturbative QCD within the {\tt xFitter} framework.  
We describe all available diffractive DIS data sets from HERA and the 
most recent high-precision H1/ZEUS combined measurements 
considering three different scenarios.
First, we extract the diffractive PDFs considering the standard twist-2 contribution. 
Then, we include the twist-4 correction from the longitudinal virtual photons.
Finally, the contribution of subleading Reggeon exchange to the 
structure-function $F_2^D$ is also examined.
For the contribution of heavy flavors, we utilize the Thorne-Roberts  
general mass variable number scheme. 
We show that for those corrections, in particular, the twist-4 contribution 
allows to include the high-$\beta$ region and leads to 
a better description of the 
diffractive DIS data sets.  
We find that the inclusion of the subleading Reggeon exchange significantly
improves the description of the diffractive DIS cross-section measurements. 
The resulting sets are in good agreement with all diffractive DIS  data analyzed, 
which cover a wider kinematical range than in previous fits. The {\tt SKMHS22} 
diffractive PDFs sets presented in this work are available 
via the {\tt LHAPDF} interface.
We also make suggestions for future research in this area. 

\end{abstract}
%

\maketitle
\tableofcontents{}

%
\section{Introduction}\label{sec:introduction}
%

One of the most important experimental findimgs reported 
by the H1 and ZEUS collaborations at the HERA collider, 
working at the center of the mass-energy of about $\sqrt{s} = 318$ GeV, is the observation of 
a significant fraction of about 8-10\% of large rapidity gap events in 
a diffractive deep inelastic scattering (DIS) 
processes~\cite{ZEUS:2008xhs,H1:2006zyl,ZEUS:2009uxs,H1:2006zxb,H1:2007oqt}.

Such diffractive DIS processes allow us to define a non-perturbative diffractive parton 
distribution functions (diffractive PDFs) that can be extracted from a QCD analysis of relevant
data~\cite{H1:2006zyl,ZEUS:2009uxs}. 
According to the factorization theorem, the diffractive cross-section can be expressed as a 
convolution of the diffractive PDFs and partonic hard scattering cross-sections of the  
subprocess which is calculable within perturbative QCD. 

The diffractive PDFs have properties very similar to the 
standard PDFs, especially they 
obey the same standard DGLAP evolution equation~\cite{Dokshitzer:1977sg,Gribov:1972ri,Lipatov:1974qm,Altarelli:1977zs}. 
However, they have an additional constraint due to the 
presence of a leading proton (LP) in the final state of diffractive 
processes, $\ell (k) + p(P) \to \ell (k^{\prime}) + p(P^{\prime}) + X(p_{X})$. 

Considering the diffractive factorization theorem, the diffractive PDFs can be extracted from 
the reduced cross-sections of inclusive diffractive DIS data by a QCD fit. 
It should be noted here that if the factorization theorem  would be violated 
in hadron-hadron scattering, then there is no universality for example for the diffractive jet 
production in hadron-hadron collisions~\cite{Collins:1992cv,Wusthoff:1999cr}. 
Starting from perturbative QCD, in the first approximation, diffractive DIS is described in the 
dipole framework and formed by the quark-antiquark (${q} \bar{{q}}$) and quark-antiquark-gluon (${q} \bar{q}$) 
system.

So far all, several groups extracted the diffractive PDFs from QCD analyses of the 
diffractive DIS data at the next-to-leading order (NLO) and next-to-next-to-leading order (NNLO) accuracy in 
perturbative QCD~\cite{Goharipour:2018yov,Khanpour:2019pzq,ZEUS:2009uxs,Ceccopieri:2016rga,H1:2006zyl,Maktoubian:2019ppi}. 
In Ref.~\cite{Goharipour:2018yov}, the authors presented the first NLO determination of the diffractive 
PDFs and their uncertainties within the {\tt xFitter} framework~\cite{Alekhin:2014irh,xFitterDevelopersTeam:2022koz}.
Ref.~\cite{Khanpour:2019pzq} presented the first NNLO determination of the diffractive PDFs,
and  the framework of fracture functions is used in the QCD
analysis~\cite{Trentadue:1993ka,deFlorian:1998rj,Ceccopieri:2016rga}. 
Some of these studies, such as ZEUS-2010-dPDFs~\cite{ZEUS:2009uxs}, also include the dijet 
cross-section measurement to determine the well-constrained gluon PDFs.

In this paper, we report a new QCD fit of diffractive PDFs 
to the HERA inclusive data 
in diffractive DIS at the NLO in perturbative QCD within the {\tt xFitter} framework~\cite{Alekhin:2014irh}.  
The present fit also includes the high-precision H1/ZEUS combined measurements of the 
diffractive DIS cross-section~\cite{H1:2012xlc}. 
The inclusion of the most recent HERA combined data, together 
with the twist-4 corrections from the longitudinal virtual photons and the contribution of 
subleading Reggeon exchanges to the structure-function, 
provide well-established diffractive PDFs sets. 
We show that these corrections, in particular, the twist-4 contribution allows us to 
explore the high-$\beta$ region, and the inclusion of the subleading Reggeons gives the 
best description of the diffractive DIS data. 
In addition, by considering such corrections, 
one could relax the kinematical cuts that one needs to apply to the data.

This paper is organized in the following way:
In Sec.~\ref{Theoretical-Framework} we discuss the theoretical 
framework of the {\tt SKMHS22} diffractive PDFs determination, 
including the computation of the diffractive DIS cross-section, 
the evolution of diffractive PDFs, and the corresponding factorization theorem.
This section also includes our choice of physical parameters and the 
heavy quark contributions to the diffractive DIS processes.
The higher twist contribution considered in {\tt SKMHS22} QCD analysis 
 also discussed in detail in this section.
In Sec.~\ref{global-QCD-analysis} we present the details of the {\tt SKMHS22} 
diffractive PDFs global QCD analysis and fitting methodology. 
Specifically, we focus on the {\tt SKMHS22} parametrization, 
the minimization strategy, and the method of uncertainty estimation. 
We also present the diffractive data set used in {\tt SKMHS22} analysis, 
along with the corresponding
observables and kinematic cuts applied to the data samples.
In Sec.~\ref{sec:results} we present in detail the {\tt SKMHS22} sets. 
The perturbative convergence 
upon inclusion of the higher twist corrections is also discussed in 
this section. 
The fit quality and the theory/data comparison are
presented and discussed in this section as well.
Finally, in Sec.~\ref{Conclusion} we summarize our findings  and 
outline possible future developments.

%
\section{Theoretical Framework}\label{Theoretical-Framework}

In the following section, we describe the standard 
theoretical framework which 
is perturbative QCD (pQCD) for the typical event 
with a large rapidity gap (LRG) 
for diffractive DIS processes.
We discuss in detail the calculation of diffractive 
DIS reduced cross-section, the relevant 
factorization theorem, and our approaches to 
consider the heavy flavors contributions. 
We also provide the details of the diffractive 
structure-function (diffractive SF) taking the 
twist-4 and Reggeon corrections into account.

%
\subsection{Diffractive DIS cross section}\label{sec:Diffractive-DIS-cross-section}

\begin{figure}[t!]
\vspace{0.20cm}
\resizebox{0.450\textwidth}{!}{\includegraphics{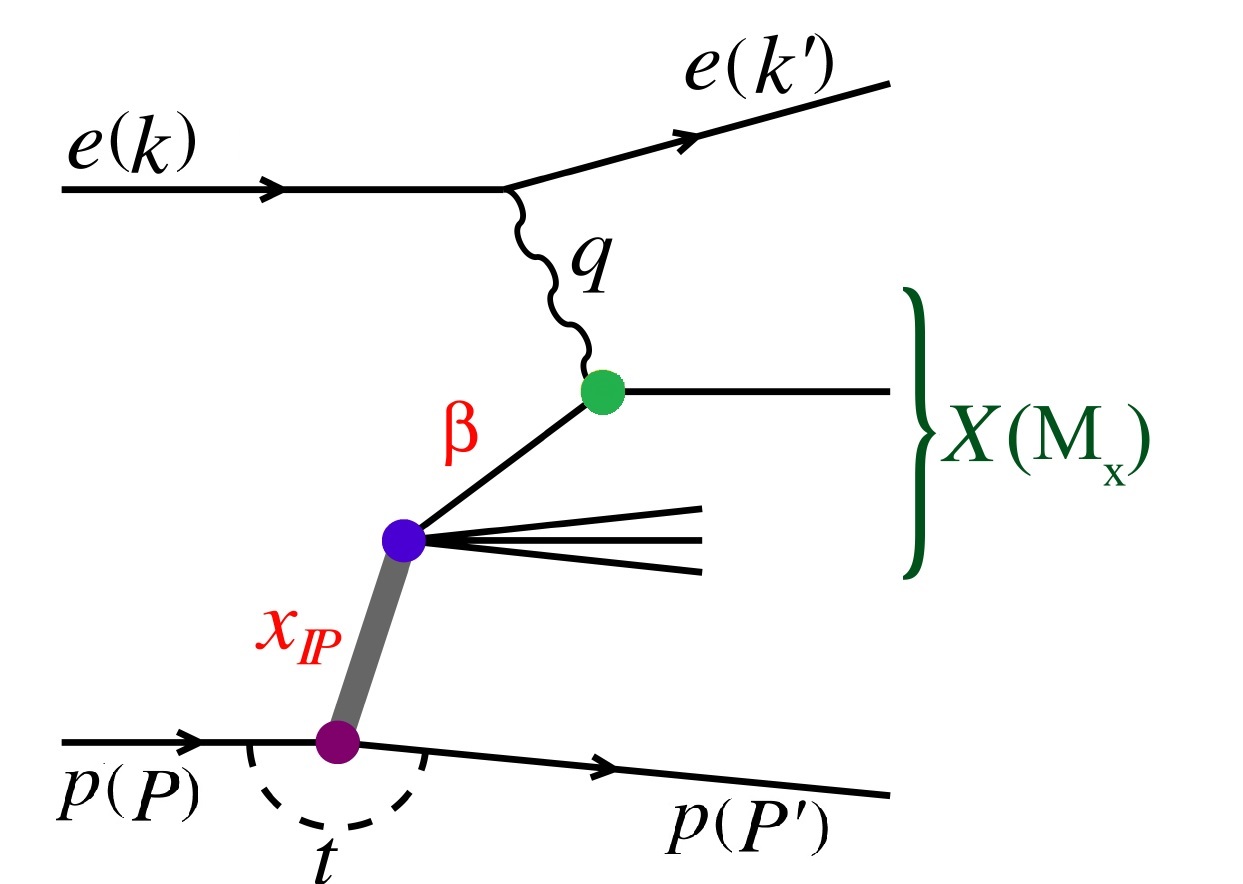}}  
\begin{center}
\caption{{\small 
Diagram for diffractive DIS $ep \rightarrow ep {X}$. 
The four-momenta are indicated as well (in teh round brackets). 
The diffractive scattered proton is distinguished 
from the diffractive mass ${X}$.} 
\label{fig:Feynman}}
\end{center}
\end{figure}

In Fig.~\ref{fig:Feynman}, we display the Feynman diagram for diffractive 
DIS in the single-photon approximation. 
In the neutral current (NC) diffractive DIS process $ep \rightarrow ep{X}$, we have the 
incoming positron or electron in the initial state in which scatters off an 
an incoming proton with the four-momentum ${k}$ and ${P}$, respectively. 
As one can see from the Feynman diagram, in the final state, 
the proton with the four-momentum of $P^\prime$ remains intact
and there is a rapidity gap between the proton in the final state and the diffractive
system $X$ and outgoing electron with four-momentum $k^\prime$. 

In order to calculate the reduced diffractive 
cross-section for such a process, 
one needs to introduce the standard set of kinematical variables 
\begin{equation}
\label{eq:kinematic-variable}
Q^{2} = -q^{2} = 
({k} - {k}^{\prime})\,, 
\quad {y} = \frac{{P.q}}{{P.k}}\,, 
\quad {x} = \frac{{-q^2}}{{2P.q}}\,,
\end{equation}
which are the photon virtuality $Q^{2}$, the inelasticity ${y}$, 
and the Bjorken variable $x$, respectively. 

In the case of diffractive DIS, one needs to introduce an 
additional kinematical variable ${\beta}$ which is 
defined to be the 
momentum fraction carried by the struck parton 
with respect to the diffractive exchange.
The kinematical variable ${\beta}$ is given by,
\begin{equation}
\label{eq:beta}
{\beta} = \frac{Q^{2}}
{{2} (P - P^{\prime})} = 
\frac{Q^{2}}
{M_X^{2} + Q^{2} - t} \,,
\end{equation}
where $M_{X}$ is the invariant mass of the diffractive final state, 
produced by the diffractive dissociation of the exchanged virtual photon, and 
the variable $t = ({P} - {P}^{\prime})$ is the squared four-momentum 
transferred at the proton vertex.

The experimental diffractive DIS data sets 
are provided by the H1 and ZEUS 
collaborations at HERA in the form of the so-called reduced cross 
section $\sigma_r^{D3}(\beta, Q^{2}; {x}_{\pom}, t)$, where $x_{\pom}$ is defined to be
the longitudinal momentum fraction 
lost by the incoming proton. 
The longitudinal momentum fraction $x_{\pom}$ 
satisfies the relation $x = {\beta} x_{\pom}$.
The $t$-integrated differential cross-section for the diffractive DIS processes can 
be written in terms of the reduced cross-section as, 
\begin{equation}
\label{eq:cross-section}
\frac{d\sigma^{e p \rightarrow e p {X}}} 
{d{\beta} dQ^{2} 
dx_{\pom}} = 
\frac{{2} {\pi}
\alpha^{2}}{{\beta} 
Q^{4}} \left[1 + (1 - {y})^{2} \right]
\sigma_r^{D3}({\beta}, Q^{2}; 
{x_{\pom}})\,.
\end{equation}

In the one-photon approximation, the reduced diffractive cross-section 
can be written in terms of two diffractive 
structure functions~\cite{H1:2006zyl,ZEUS:2009uxs,Goharipour:2018yov}. 
This reads,
\begin{align}
\label{eq:reduced cross-section}
\sigma_r^{D(3)}({\beta}, 
Q^2; x_{\pom})=& 
F_2^{D(3)}({\beta}, {Q}^{2}; 
{x_{\pom}}) 
\\ \nonumber
& -\frac{y^{2}}{1 + (1 - {y})^{2}} 
F_{L}^{D(3)}
({\beta}, {Q}^{2}; {x_{\pom}}).
\end{align}
It should be noted here that for the ${y}$ not close to the unity, the
contribution of the longitudinal structure function $F_{L}^{D(3)}$
to the reduced cross-sections can be neglected. 
Since we use the diffractive DIS data sets at HERA for the 
reduced cross-section, we follow the recent study by 
{\tt GKG18}~\cite{Goharipour:2018yov} 
and consider this contribution in our QCD analysis.
%
%
%
%

%
\subsection{Factorization theorem for the diffractive DIS}\label{Factorization theorem }
%

The main idea of diffractive DIS was proposed 
for the first time by Ingelman and Schlein~\cite{Ingelman:1984ns}. 
According to the Ingelman-Schlein (IS) model, 
the diffractive processes in DIS are interpreted 
in terms of the exchange of the leading Regge trajectory. 
The diffractive process includes two steps which are the 
emission of the pomeron from a 
proton and subsequent hard scattering of a virtual photon 
on partons in the pomeron. 
Therefore the pomeron is considered to have a partonic structure 
as do  hadrons.
Hence,  the diffractive 
structure functions factorizes into a pomeron flux and a 
pomeron structure function~\cite{Collins:1996fb,Collins:2001ga,Collins:1997sr}.  
In analogy to the inclusive DIS, the diffractive structure functions 
can be written as a 
convolution of the non-perturbative diffractive PDFs which satisfy the standard  
DGLAP evolution equations~\cite{Dokshitzer:1977sg,Gribov:1972ri,Lipatov:1974qm,Altarelli:1977zs}, 
and the hard scattering coefficient functions.  
It is given by,
\begin{align}
\label{eq:diffractive-structure-function}
F&_{2/L}^{D(4)}({\beta}, Q^{2}; 
x_{\pom}, {t})=  \nonumber \\
&\sum_{i} \int_{\beta}^{1} 
\frac{dz}{z} {C}_{{2}/{L}, i} 
\left(\frac{{\beta}}{z}\right)
{f}_{i}^{D}(z, Q^{2}; {x}_{\pom}, {t}),
\end{align}
where the sum runs over all parton flavors (gluon, $d$-quark, $u$-quark, etc.).

Here the long-distance quantity  $f_{i}^D(z, Q^{2}; x_{\pom}, {t})$ denotes 
the non-perturbative part which can be determined by a 
global QCD analysis of 
the available diffractive experimental data sets. 
The Wilson coefficient functions $C_{{2}/{L}, {i}}$  
in the above equation 
describe the hard scattering of the virtual
photon on a parton ${i}$ and are the same as the coefficient
functions known from the 
inclusive DIS and calculable in perturbative QCD~\cite{Vermaseren:2005qc}.  
It has been shown that the description of the experimental data is  very 
good when factorization is assumed~\cite{H1:2006zyl,ZEUS:2009uxs,Goharipour:2018yov}. 
The vertex factorization states that the diffractive PDFs should be 
factorized into the product of two terms, 
one of them depends on the $x_{\pom}$ and $t$, and the 
another one is a function of ${\beta}$ and $Q^{2}$.
Hence, the diffractive PDFs $f_{i/p}^D({\beta}, Q^{2}; x_{\pom}, {t})$ is given by,  
\begin{align}
\label{eq:DPDFs}
f_{i/p}^D({\beta}, Q^{2}; x_{\pom}, {t})= & 
f_{\pom/p}(x_{\pom}, {t})
f_{i/\pom}({\beta}, Q^{2}) \nonumber\\
+&f_{\reg/p}(x_{\pom}, {t}) 
f_{i/\reg}^{\reg}({\beta}, Q^{2}),
\end{align}
where $f_{\pom/p}(x_{\pom}, {t})$ and $f_{\reg/p}(x_{\pom}, {t})$ are the Pomeron and 
Reggeon flux-factors, respectively. 
These describe the emission of the 
Pomeron and Reggeon from the proton target. 
The Pomeron and Reggeon partonic structures given
by the parton distributions 
$f_{{i}/\pom}({\beta}, {Q}^{2})$ and $f_{i/\reg}^{\reg}({\beta}, Q^{2})$. 
The  parametrization and determination of these functions will be 
discussed in detail in section~\ref{global-QCD-analysis}.

%
\subsection{Heavy flavour contributions}\label{subsec:Heavy flavour}
%

The study of the heavy quark flavor contributions to the DIS processes enables us to precisely test QCD 
and the strong interactions. 
In this respect, their contributions have an important impact on 
the PDFs~\cite{Ball:2021leu,Boroun:2021ckh,Harland-Lang:2014zoa,Ball:2022hsh}, 
FFs~\cite{Salajegheh:2019nea,Salajegheh:2019ach,Soleymaninia:2022qjf}, 
and diffractive PDFs~\cite{Maktoubian:2019ppi,Khanpour:2019pzq,Goharipour:2018yov}
extracted from the global QCD analysis. 

Generally speaking, there are two regimes for the treatment of 
heavy quark production.  
The first region is $Q^{2} \sim m_{h}^{2}$ 
where $m_{h}$ is the heavy quark mass. 
The massive quarks are produced
in the final state and they are not treated as an active parton within the nucleon. 
This regime is interpreted using the ``Fixed Flavour Number Scheme'' (FFNS). 
In this scheme, the light quarks are considered to be the 
active partons inside the nucleon, 
and the number of flavors needs to be fixed to  $n_{f} = 3$.  
The FFNS is not accurate and reliable for scales much 
greater than the heavy quark mass threshold $m_{h}^{2}$.
At higher energy scales, ${Q}^{2} \gg m_{h}^{2}$, the heavy quarks behave as 
massless partons within the hadron. 
It that case, logarithmic terms $\sim Q^{2}/m_h^{2}$ 
are automatically summed through the solutions of 
the DGLAP evolution equations for the heavy quark distributions. 
The simplest approach which describes the ${Q}^{2}/m_h^{2} \to \infty$ 
limit is the ``Zero Mass Variable Flavor Number Scheme'' (ZM-VNS), 
which ignores all the $\mathcal{O}(m_{h}^{2}/{Q}^{2})$ corrections. 
In summary, at very large scales, where the resummation of large 
logarithms are pertinent, the ZM-VFNS would be more precise, and in 
the limits close to the heavy quark mass threshold ${m}_{h}$, the 
FFNS works well enough. 
In order to present the correct scheme and 
to obtain the best description of these two 
limits of $Q^2\leq m_h^2$ and $Q^2\gg m_h^2$, one needs to take into 
account the  
``General Mass Variable Flavor Number Scheme'' (GM-VFNS)~\cite{Thorne:2012az}. 
In this scheme, the DIS structure function can be written as follows,  
\begin{equation}
\label{eq:GM-VFNS}
F(x, Q^{2}) = C_{j}^{{\text {GMVFN}}, n_{f} + m}
(Q^{2}/m_{h}^{2}) \otimes f_{j}^{n_{f} + {m}}(Q^{2}),
\end{equation}
where ${n}_{f}$ is the number of active light quark 
flavors and ${m}$ is the number of heavy quarks.
Unlike the ZM-VFNS, the hard scattering coefficient 
functions ${C}_{k}^{{\text {FF}}, n_{f}}$ 
depend on the $(Q^2/m_h^2)$ but reduce to the 
zero mass approach as $Q^{2}/m_h^{2} \to \infty$. 
Considering the transition from ${n}_{f}$ active quarks to ${n}_{f} + 1$ ones, 
one could write, 
\begin{align}
\label{eq:GM-VFNS1}
F(x, {Q}^{2})=&{C}_{j}^{{\text {GMVFN}}, 
n_{f} + 1}(Q^{2}/m_{h}^{2}) 
\otimes f_{j}^{n_{f} + 1}(Q^{2})  \nonumber \\
=&C_{j}^{{\text {GMVFN}}, 
n_{f} + {1}}(Q^{2}/m_{h}^{2}) 
\otimes A_{jk}(Q^{2}/m_{h}^{2}) \nonumber \\
& \qquad\qquad \qquad 
\qquad \quad \quad \otimes 
{f}_{k}^{n_{f}}(Q^{2}) \nonumber\\ \equiv&  
C_k^{FF, {n}_{f}}(Q^{2}/{m}_{h}^{2}) \otimes 
f_{k}^{n_{f}}(Q^{2}),
\end{align}
where the matrix elements $A_{jk}(Q^{2}/m_{h}^{2})$ are 
calculated and presented in Ref.~\cite{Buza:1996wv}. 
The $C_{k}^{{\text {FF}}, {n}_{f}}(Q^2/m_{h}^{2})$ coefficient in 
the above equation is given by,
\begin{equation}
\label{eq:GM-VFNS2}
C_{k}^{{\text {FF}}, 
n_{f}}(Q^{2}/m_{h}^{2}) 
\equiv A_{jk}(Q^2/m_{h}^{2}) 
\otimes f_{k}^{n_f}(Q^{2}).
\end{equation}
As we mentioned before, the coefficient functions 
must behave towards the 
massless limit as ${Q}^{2}/m_h^{2} \to {\infty}$, 
as given by Eq.~\eqref{eq:GM-VFNS2}.
The analysis presented in this work 
is based on the Thorne and Roberts (TR) GM-VFNS which 
covers smoothly from FFNS scheme at low ${Q}^{2}$ to 
the ZM-VFNS descritption at high energy scale ${Q}^{2}$, 
and in this regard, our choice gives the best description of the heavy 
flavor effects on  the diffractive structure functions. 
In this study, following the MMHT14 collaboration, 
we adopt the values for heavy 
quarks as ${m}_{c} = 1.40\,$GeV and ${m}_{b} = 4.75\,$GeV~\cite{Harland-Lang:2015qea}. 
The strong coupling constant is fixed to 
${\alpha}_s(M_Z^2) = 0.1185$~\cite{ParticleDataGroup:2018ovx}.

%
\subsection{Twist-4 correction}\label{subsec:twist4}

In this section, we describe in detail the twist-4 correction considered in 
the {\tt SKMHS22} QCD analysis.
Generally speaking, the higher twist contribution is 
proportional to the 
terms depending on  ${1}/Q^{2}$, and hence, would be highly 
suppressed at a high-energy scale ${Q}^{2}$ in the DIS process. 
Nonetheless, in the color dipole framework this term for 
${M} \rightarrow 0$ or 
${\beta} \rightarrow {1}$ dominates over the 
twist-2 contribution. 
Hence, the dipole picture provides a powerful 
framework in which the QCD-based saturation
models can be used to investigate the diffractive DIS data.

For interaction of the color dipole with a proton, 
one could consider two different 
models which are the quark dipole (${q} \bar{q}$ system) 
and gluon dipole (${q} \bar{q}g$) system. 
We refer the reader to the Ref.~\cite{Golec-Biernat:2008mko} for 
a clear review.

As we mentioned earlier, 
for the diffractive cross-section, the virtual photon could be 
transversely or longitudinally polarized. 
Therefore, according to the different polarization, one can 
decompose it into a transverse and a longitudinal part. 
It turns out for the leading order in ${Q}^{2}$ for ${\beta} \rightarrow {1}$, 
the ${q} \bar{q}$ and  ${q} \bar{q}g$ from transverse virtual photons vanish 
proportional to the $({1} - {\beta})$,
whereas the longitudinal part which is of higher twist 
and gives a finite contribution. 
In  diffractive DIS, for ${M} \ll {Q}^{2}$, the ${q} \bar{q}$ 
contribution to the final state 
dominates over ${q} \bar{q} {g}$ of the photon wave function. 
Thus, the ${L}{q} \bar{q}$ component is not negligible
at higher value of ${\beta}$ and 
has an important contribution in 
the longitudinal diffractive structure function.  
The ${F}_{Lq \bar{q}}^{D}$ can be written in term of the 
Bessel functions ${J}_{0}$ and ${K}_{0}$ and 
dipole cross section 
$\hat{\sigma}({x}_{\pom}, {r})$~\cite{Golec-Biernat:2007mao,Golec-Biernat:2008mko},
\begin{align}
\label{twist-4}
F_{Lq\bar{q}}^{D} = 
&\frac{3}
{16 {\pi}^{4} x_{\pom}} 
e^{-B_{D} \vert t \vert}\sum_{f} 
e_{f}^{2}\frac{{\beta}^3} 
{({1}-{\beta})^{4}} \nonumber \\
&\times \int_{0}^{[Q^2(1 - 
{\beta})]/{4} {{\beta}}} dk^{2} 
\frac{k^{2}/Q^{2}} 
{\sqrt{{1} - \frac{{4} {\beta}}
{{1} - {\beta}}
\frac{k^{2}}{Q^2}}} \nonumber \\
&\times \left( k^{2} \int_{0}^{\infty} dr 
r {K}_{0} 
\left(\sqrt{\frac{{\beta}}{{1} - 
{\beta}}}kr\right) 
{J}_{0}(kr) \hat{\sigma}({x}_{\pom}, 
{r}) \right)^{2}\,.
\end{align}

The main idea for the dipole  approach is that the 
photon splits up into a quark-antiquark
pair (dipole), 
which then scatters on the proton target. 
Following the studies presented in Refs.~\cite{Golec-Biernat:1998zce}, 
the dipole cross-section is considered to have the 
following simple form  in our QCD analysis,
\begin{equation}
\hat{\sigma}({x}_{\pom}, {r}) = 
\sigma_{0} \{{1} - 
\exp(-r^{2} {Q}_{s}^{2}/{4})\},
\end{equation}
where ${r}$ indicates the separation between
the quark and the antiquark. 
The saturation momentum $Q_{s}^{2}=(x_{\pom}/x_{0})^{-\lambda}$ GeV$^{2}$ 
is responsible for the transition to the saturation regime. 
The parameters $\sigma_{0} = 29$~mb, $x_{0} = 4\times10^{-5}$, and 
$\lambda = 0.28$ are taken from Ref.~\cite{Golec-Biernat:1998zce}. 
This definition of the dipole-proton 
cross-section presents a 
good description of the inclusive HERA diffractive DIS data sets.
Our QCD analysis which includes the twist-4 correction is 
called  {\tt SKMHS22-tw2-tw4}.
The effect of such corrections on the extracted diffractive 
PDFs is  discussed in section.~\ref{sec:results}.

%
\subsection{Reggeon contribution}\label{subsec:Reggeon}

For the higher value of ${x}_{\pom}$, these diffractive 
DIS data sets include the 
contributions which decrease with energy.
In order to truly describe this effect one can include the 
contributions of the 
subleading Reggeon which breaks the factorization of the 
diffractive structure function. 
In order to take into account such contributions, 
we add the following Reggeon 
contribution to the diffractive 
structure function~$F^2_D$~\cite{Golec-Biernat:1996bax,Golec-Biernat:1997vbr}
\begin{equation}\label{eq:ReggeonSF}
\frac{dF_{2}^{R}}{d {x}_{\pom} dt}
({x}, Q^{2}, x_{\pom}, {t}) = 
f^R(x_{\pom}, {t}) \, 
F_{2}^{R}({\beta}, {Q}^{2})\,,
\end{equation}
where the ${f}^{R}(x_{\pom}, {t})$ is the Reggeon flux, 
and the ${F}_{2}^{R}({\beta}, Q^{2})$ 
is the Reggeon structure function. 
In principle, we
should consider different Regge pole 
contributions and sum over them and
include interference terms. 
Approximately, one can neglect the interference
terms between Reggeons and the Pomeron and also 
between different Reggeons. 
Therefore for the Reggeon flux, 
we consider the following formulas:
\begin{equation}
\label{eq:Reggeonsum}
f^{R}({x}_{\pom}, {t})  = 
\sum_{R_{i}} f^{R_{i}}
(x_{\pom}, {t})\,,
\end{equation}
and
\begin{equation}\label{eq:Reggeonflux}
f^{R_{i}}(x_{\pom}, {t})=
\frac{F^{2}_{i}(0)} {{8} \pi} 
e^{-\vert t 
\vert /{\lambda}_{i}^{2}} 
{C}_{i}(t) {x}_{\pom}^{{1} - {2} 
\alpha_{i}(t)}\,.
\end{equation}
Here, $C_i(t) = {4} \cos^2[\pi {\alpha}_i(t)/2]$ and 
$C_i(t) = {4} \sin^2[\pi{\alpha}_i(t)/2]$
are the signature factors for the even 
Reggeon ($f_{2}$) and the odd Reggeon (${\omega}$), respectively. 
The Reggeon trajectory is given by 
${\alpha}_{i}(t) = 0.5475 + (1.0$ GeV$^{-2})t$. 
From Ref.~\cite{Golec-Biernat:1997vbr}, 
$F^{2}_{f_{2}}(0) = 194$~GeV$^{-2}$ and  $F^{2}_{\omega}(0)=52$~GeV$^{-2}$ 
which denote the Reggeon couplings to the proton. 
$\lambda_{i}^{2} = 0.65$~GeV is known from  Reggeon phenomenology 
in hadronic reactions. 
The analysis for the isoscalar Reggeons $f_{2}, {\omega}$ 
shows that the Reggeon contribution 
to the diffractive structure-function 
becomes important for ${x}_{\pom} > 0.01$~\cite{H1:2006uea}. 
Our assumption for the Reggeon structure functions is 
that they are the same for all Reggeons and 
${F}_{2}^{R}({\beta}, {Q}^{2})$ is related
to the parton distributions 
in the Reggeons in a conventional way 
and can be determined by the fit to diffractive DIS
data from HERA~\cite{H1:2012xlc,H1:2012pbl,H1:2011jpo}. 
For the Reggeon structure-function, we 
consider the following parametrization form,
\begin{equation}
\label{eq:Reggeonstructure}
{F}_{2}^{R}({\beta}) = w_{1} {\beta}^{w_{2}}
(1 - {\beta})^{w_3} (1 + w_4 \sqrt{{\beta}} + w_{5} {\beta}^2)~,
\end{equation}
where $w_1$, $w_2$, $w_3$ and $w_4$ are free fit parameters 
and they need  to be determined from a global QCD analysis.
The parameter $w_2$ controls the shape of $F_2^R({\beta})$ in the low-$\beta$ 
region, while  $w_3$ controls the high-$\beta$ region. 
The parameters $w_4$ and $w_5$ are 
considered to be fixed to zero, as the current
diffractive DIS data sets do not have enough power 
to constrain all the shape parameters.
Our QCD analysis which includes the Reggeon contribution is 
called  {\tt SKMHS22-tw2-tw4-RC}, and
the effect arising from the inclusion of such correction is 
discussed in section.~\ref{sec:results}.

%
\section{Diffractive PDFs global QCD analysis details}\label{global-QCD-analysis} 

The analysis of diffractive PDFs is a QCD optimization problem  and 
can be seen as having four  elements to it. 
The first and most important element 
is of course the experimental data and their uncertainty. 
The second  element is a theoretical model used to  describe
these phenomenon. For the analysis of non-perturbative 
objects such as diffractive PDFs this theoretical 
framework can have some adjustable parameters as well as 
a number of constraints (say for ensuring of factorization). 
Another part of the problem is the objective function 
which we wish to minimize/maximize.  
Here, this function
is a $\chi^{2}$ function that is defined by 
the theoretical framework and includes the data uncertainty. 
The last part of the problem is the optimization 
method which is the essential step to finding the solution.

In this section, we present the methodology of 
our global QCD analysis to obtain the diffractive 
PDFs and their uncertainties at NLO accuracy in perturbative QCD. 
We describe the details of the data sets 
used for the following fit 
in section~\ref{subsec:data} , the parametrization of 
diffractive PDFs is discussed in section~\ref{subsec:param}, 
and the methodology of minimization and uncertainties of 
our diffractive PDFs are described in section~\ref{subsec:uncertainies}.

%
\subsection{Experimental data sets}\label{subsec:data}

\begin{table*}
\caption{\small List of all diffractive DIS data points with their properties 
used in the {\tt SKMHS22}  global QCD analysis. 
For each data set we  provide the 
kinematical coverage of ${\beta}$, ${{x}_{\pom}}$, and ${Q}^{2}$. The number of data points is 
displayed as well. 
The details of the kinematical cuts applied on these data sets are explained in the text. } 
\label{tab:DDISdata}
\begin{tabular}{l | c  c  c  c  c  c }
\hline\hline
{\text{Experiment}} & {\text{Observable}} & [$\beta^{{\text{min}}}, {\beta}^{{\text{max}}}$] & [${x_{\pom}}^{\rm{min}}, {x_{\pom}}^{\rm{max}}$]  & ${Q}^{2}\,[{\text{GeV}}^2]$  & \# of points & {\text{Reference}}
\tabularnewline
\hline\hline
{\text{H1-LRG-11}} $\sqrt{s} = {225}$ GeV & ${\sigma}_{r}^{D(3)}$ & [$0.033$--$0.88$]    & [$5{\times} 10^{-4}$ -- $3{\times} 10^{-3}$] & 4--44 & \textbf{22}  & \cite{H1:2011jpo} 
 \\
{\text{H1-LRG-11}} $\sqrt{s} = {252}$ GeV & ${\sigma}_{r}^{D(3)}$ & [$0.033$--$0.88$]    & [$5{\times} 10^{-4}$ -- $3{\times} 10^{-3}$] & 4--44 & \textbf{21}  &  \cite{H1:2011jpo} 
\\
{\text{H1-LRG-11}} $\sqrt{s} = {319}$ GeV & ${\sigma}_{r}^{D(3)}$ & [$0.089$--$0.88$]    & [$5{\times} 10^{-4}$ -- $3{\times} 10^{-3}$] & 11.5--44 & \textbf{14} &  \cite{H1:2011jpo}   
\\	
{\text{H1-LRG-12}} & ${\sigma}_{r}^{D(3)}$ & [$0.0017$--$0.80$]   & [$3{\times} 10^{-4}$ -- $3{\times} 10^{-2}$] & 3.5--1600 & \textbf{277}   &  \cite{H1:2012pbl} 
\\	
{\text{H1/ZEUS combined}} & ${\sigma}_{r}^{D(3)}$  &   [$0.0018$--$0.816$]   & [$3{\times} 10^{-4}$ -- $9 {\times} 10^{-2}$] & 2.5--200 & \textbf{192}  & \cite{H1:2012xlc}   
\\			
\hline \hline
\multicolumn{1}{c}{\textbf{Total data}} ~~&~~ &~~ &~~& ~~&~~\textbf{526}  \\  \hline  \hline
\end{tabular}
\end{table*}

In this section, we describe all available 
inclusive diffractive DIS data sets in detail. 
However, one needs to
apply some kinematical cuts in order to avoid 
nonperturbative effects 
and ensure that
only the data sets for which the 
available perturbative QCD
treatment is adequate are included in the QCD analysis. 
We attempt to include more data sets, and hopefully by including the twist-4 
and Reggeon contributions, one 
could relax some kinematical cuts 
that need to be applied to the data.

\begin{figure*}
\vspace{0.20cm}
\resizebox{0.90\textwidth}{!}{\includegraphics{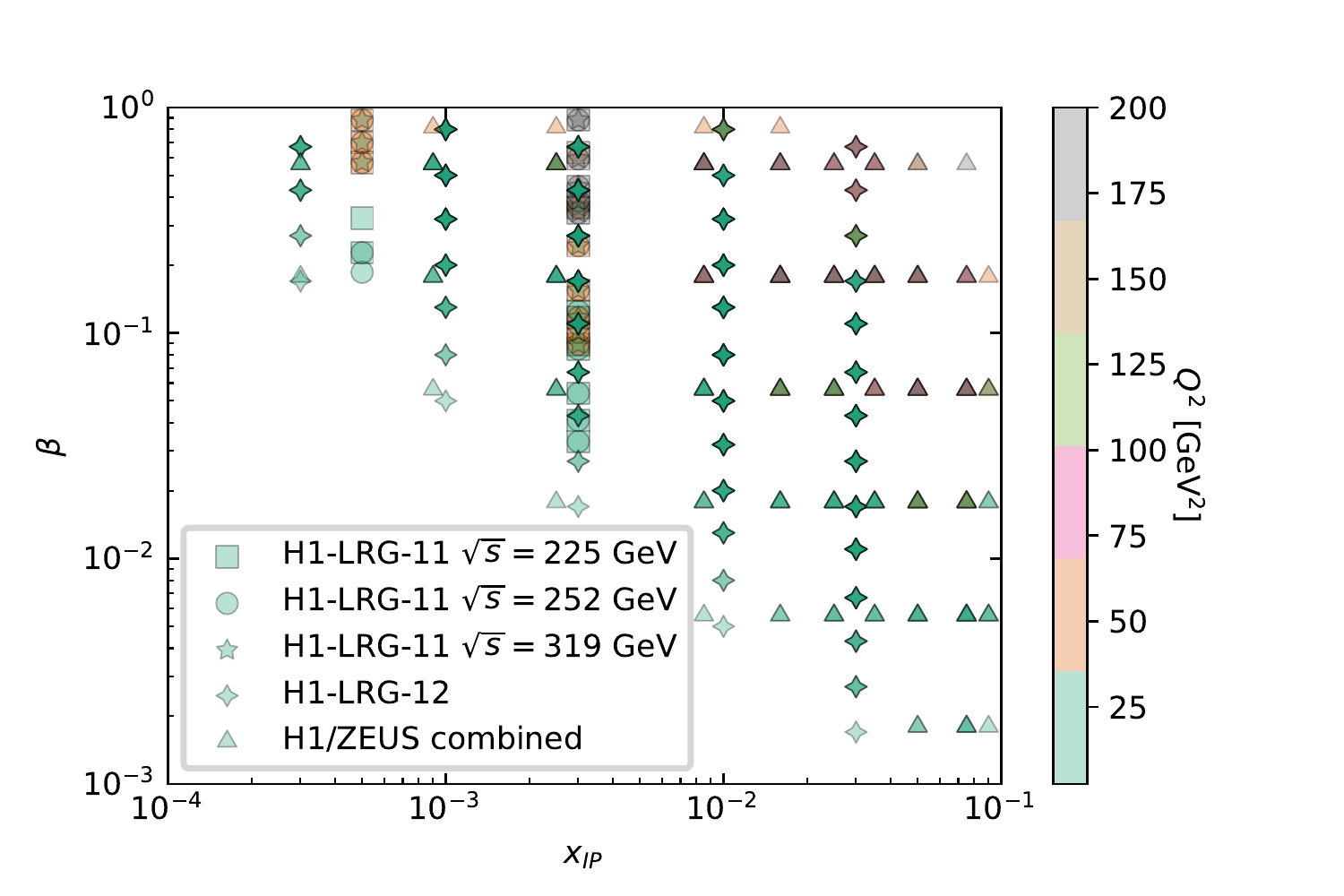}}
\begin{center}
\caption{{\small
The kinematic coverage in the ($\beta$, ${x_{\pom}}$, Q$^2$) plane of the 
{\tt SKMHS22} data set. 
The data points are classified by different experiments at
HERA. 
} 
\label{fig:databetaxQ2}}
\end{center}
\end{figure*}

The kinematic coverage in the (${\beta}$, ${x_{\pom}}$, {Q}$^2$) plane of 
the complete {\tt SKMHS22} data set is 
displayed in Fig.~\ref{fig:databetaxQ2}. 
The data points are classified by different experiments at HERA.
As customary, some kinematic cuts need to be applied to 
the diffractive DIS cross-section measurements.
%
%
%
%
The list of diffractive DIS data sets and their 
properties used in the {\tt SKMHS22}  global 
analysis are shown in Table~\ref{tab:DDISdata}. 
All the measurements are presented in terms of the $t$-integrated reduced 
diffractive DIS cross-section  measurements ${D}_{r}^{D(3)}(ep \to ep{X})$. 
In the {\tt SKMHS22} QCD analysis, 
we also analyze the combined measurement of 
the inclusive diffractive cross-section 
presented by the H1 and ZEUS Collaborations at HERA (H1/ZEUS combined)~\cite{H1:2012xlc}. 
They used samples of diffractive DIS $ep$ data at the center-of-mass energy of $\sqrt{s} = {318}$~GeV at 
the HERA collider where the leading protons are detected by 
appropriative spectrometers. 
This high-precision measurement combined all 
the previous H1 FPS HERA I~\cite{H1:2006uea}, 
H1 FPS HERA II~\cite{Aaron:2010aa}, 
ZEUS LPS 1~\cite{ZEUS:2004luu}, and 
ZEUS LPS 2~\cite{ZEUS:2008xhs} data sets. 
These measurements cover the photon virtuality interval $2.5 <{Q}^{2}< 200$~GeV$^{2}$, 
$3.5 {\times} 10^{-4}< {x_{\pom}} < 0.09$ in proton fractional momentum loss, 
$0.09< |t| < 0.55$~Gev$^{2}$ in squared four-momentum 
transfer at the proton
vertex and $1.8 {\times} 10^{-3} < {\beta} < 0.816$. 
We should highlight here that all H1-LRG data are 
published for the range of $\vert t \vert<1$~GeV$^{2}$, 
while the recent combined H1/ZEUS data sets are restricted 
to the range of $0.09 < \vert t \vert < 0.55$~GeV$^{2}$, and hence, 
one needs to use a global normalization factor to  
extrapolate from $0.09 < \vert t \vert < 0.55$~GeV$^{2}$ to $\vert t \vert<1$~GeV$^{2}$~\cite{Goharipour:2018yov}. 
Therefore, the combined H1/ZEUS data are corrected for the region of $\vert t \vert<1$~GeV$^{2}$.

Another data set that we have used in our QCD analysis is the 
Large Rapidity Gap (LRG) data from H1-LRG-11, 
which was measured by the H1 detector in 2006 and 2007. 
These data are derived for  three different center of mass 
energies of $\sqrt{s}={225}$, ${252}$ and ${319}$~GeV~\cite{H1:2011jpo}. 
The H1 collaboration measured the reduced cross-section in the 
photon virtualities range $4\,$GeV$^{2} {\leq} Q^{2} {\leq} 44\,$GeV$^{2}$ for 
the center of mass  
$\sqrt{s}={225}$, ${252}$ GeV, and $11.5$ GeV$^{2} {\leq} Q^2 {\leq} 44$ 
GeV$^2$ for the center-of-mass
energy of  $\sqrt{s}={319}$ GeV. 
The diffractive final state masses and the proton vertex are in the range of 
$1.25<{M}_{X}<10.84$~GeV and $\vert t \vert<1$~GeV$^{2}$, respectively. 
The diffractive variables are considered in the 
range of $5 {\times} 10^{-4} < {x_{\pom}} < 3 {\times} 10^{-3}$, 
$0.033 < {\beta} < 0.88$ for $\sqrt{s}=225$, $252$~GeV, and 
$0.089< {\beta} <0.88$ for $\sqrt{s}=319$~GeV. 

Finally the last data set that we have employed 
is the H1-LRG-12~\cite{H1:2012pbl}. 
These data for the process $e p \rightarrow e{X}{Y}$ have been 
derived by the H1 experiment at HERA.
H1-LRG-12 data covers the range of 
$3.5 < {Q}^{2} < 1600\,$GeV$^{2}$, $0.0003 \leq {x_{\pom}} \leq 0.03$, 
and $0.0017 {\leq} {\beta} {\leq} 0.8$.

We applied some kinematical cuts on  ${\beta}$, $M_{X}$, and $Q^{2}$. 
The kinematical cuts we applied are similar to those used in 
Refs.~\cite{Goharipour:2018yov,ZEUS:2009uxs}, except 
for the case of ${\beta}$. 
By considering the twist-4 and Reggeon corrections, 
we can relax the cuts to include more 
data sets than those have been used in Ref.~\cite{Goharipour:2018yov}. 
For the extraction of diffractive PDFs we 
apply $\beta \leq 0.90$ and ${M}_{X}<2$ GeV over all data sets 
used in this analysis. 
The sensitivity of data sets to the ${Q}^2$ has been tested 
in Refs.~\cite{Goharipour:2018yov,ZEUS:2009uxs}. 
The authors  finalized the cut on $Q^{2}$ by making a $\chi^2$ scan. 
In Ref.~\cite{Maktoubian:2019ppi} the authors used standard higher 
twist for structure functions and 
they obtained the best $\chi^2$ by considering the $Q^2_{\text{min}}= 6.5$~GeV$^2$, and in 
Ref.~\cite{Goharipour:2018yov} to extract the diffractive PDFs 
they found the best cut is $Q^2_{min}=9$~GeV$^2$. 
In this work and after testing the results for the $\chi^2$ we found the best 
agreement of theory and data will be 
achieved by taking $Q^2_{min}=9$~GeV$^2$.  
After applying these kinematical cuts the total 
number of data points is reduced to the 302.

%
\subsection{{\tt SKMHS22} Diffractive PDFs parametrization form}\label{subsec:param}

Diffractive PDFs are nonperturbative quantities and 
cannot be calculated in perturbative QCD.
Therefore, for their functional dependence a parametric 
form with some unknown parameters should 
be considered at the input scale.
Due the lack of diffractive DIS experimental data, 
for the quark 
distributions we 
consider all light quarks and anti-quarks densities to be equal,
${f}_{u} = {f}_{d} = {f}_{s} = f_{\bar{u}} = f_{\bar{d}} = f_{\bar{s}}$. 
The scale dependence of the 
quarks and gluon distributions ${f}_{q, g}({\beta}, {Q}^{2})$ needs to
be determined by  
the standard DGLAP evolution equations. 
We fit the diffractive quark and gluon 
distributions at the starting scale ${Q}_{0}^{2} = {1.8}$~GeV$^2$ which is 
below the charm threshold ($m_c^2=1.96$~GeV$^2$).
Since the diffractive DIS data sets can only constrain 
the sum of the diffractive 
PDFs and due to the small amount of the inclusive diffractive 
DIS data sets, one needs to
consider the less  flexible parametrization form for diffractive PDFs. 

We fit the diffractive parton distribution function at the initial scale
$Q_0^2 = 1.8$ GeV$^2$ with the following pomeron 
parton distributions which have been used in several 
analysis~\cite{Goharipour:2018yov,Maktoubian:2019ppi,H1:2006zyl}:
\begin{align}
zf_{q}(z, {Q}_{0}^{2}) = & 
{\alpha}_qz^{{\beta}_q}({1} - {z})^
{{\gamma}_q}({1} + {\eta}_q\sqrt{z} + {\xi}_q z^2), 
\label{eq:parametrizationformq} \\
zf_{g}(z, {Q}_0^2) = & \alpha_gz^{{\beta}_{g}}(1 - {z})^
{{\gamma}_g}({1} + {\eta}_g\sqrt{z} + {\xi}_{g} z^2), 
\label{eq:parametrizationformg} 
\end{align}
where ${z}$ in above equation is the longitudinal momentum 
fraction of struck parton 
with respect to the diffractive exchange. 
At the lowest order, $z = {\beta}$ but 
by including the higher orders this parameter differs from ${\beta}$ 
and this  leads to the ${0} < {\beta} < {z}$. 
In order to let the parameters ${\gamma}_q$ and ${\gamma}_g$ have enough 
freedom to achieve negative or positive values in the QCD fit we follow the
QCD analyses avalible in the litrature and add 
an additional term ${e}^{- \frac{0.001}{{1} - {z}}}$ to ensure that 
the distributions vanish for ${z} \rightarrow {1}$. 
It turned out that four parameters ${\eta}_q$, ${\eta}_g$, ${\xi}_q$ and ${\xi}_g$ 
are not well 
constrained by the experimental data, and hence, 
one needs to set them to zero. 
The heavy flavors are generated through the DGLAP 
evolution equations at the scale ${Q}^{2} > m_{c, b}$. 
As we mentioned, to consider the contributions of 
heavy flavors, we apply the Thorne-Roberts scheme GM-VFN scheme. 

The $x_{\pom}$ dependence of the diffractive PDFs 
$f_i^D({\beta}, Q^{2}; x_{\pom}, {t})$ in Eq.~\ref{eq:DPDFs}
is determined by the Pomeron and Reggeon flux with  linear 
trajectories, $\alpha_{{\pom}, {\reg}}(0) + \alpha^\prime_{{\pom}, 
{\reg}}t$~\cite{Goharipour:2018yov,Maktoubian:2019ppi,H1:2006zyl}
\begin{equation}
\label{eq:pomeronReggeonflux}
{f}_{{\pom}, {\reg}}({x}_{\pom}, t) = 
{A}_{{\pom}, {\reg}}
\frac{e^{B_{{\pom}, {\reg}}}t} {{x}_{\pom}^
{{2} \alpha_{{\pom}, {\reg}}(t)-{1}}},
\end{equation}
where the normalization factor of Reggeon ${A}_{\reg}$ and the 
Pomeron and Reggeon intercepts, ${\alpha}_{\pom}(0)$ and ${\alpha}_{\reg}(0)$ are 
free  parameters and 
will be determined from fit to the diffractive DIS data. 
We should note here that, according to Eq~\eqref{eq:DPDFs} 
the parameter ${A}_{\pom}$ is absorbed into the ${\alpha}_{q}$ and ${\alpha}_{g}$. 
The remaining parameter appearing in Eq.~\eqref{eq:pomeronReggeonflux} are 
taken from~\cite{Goharipour:2018yov}. 
In general we have twelve free fit parameters: 
six from Eq.~\eqref{eq:parametrizationformq} and \eqref{eq:parametrizationformg}, 
three form Eq.~\eqref{eq:pomeronReggeonflux}, and 
three from the Reggeon structure function in Eq.~\eqref{eq:Reggeonstructure}.

%
%
\subsection{Minimization and diffractive PDF uncertainty method}\label{subsec:uncertainies}

In this section, we  present two important parts of the  {\tt SKMHS22}  QCD analysis. 
First, we discuss the optimization method, and then we show how 
to take into account  
the data uncertainty in the final results. 
In order to estimate the free parameters of 
the diffractive PDFs given the 
experimental data of section~\ref{subsec:data} 
we apply the maximum log-likelihood method. 
If we assume that the data arise from a 
Gaussian distribution as is usually done, 
this method coincides with minimizing the $\chi^2$ estimator. 
Here we adopt the following form for the $\chi^{2}$ function~\cite{H1:2012qti,Goharipour:2018yov},
\begin{eqnarray}
\label{eq:chi2}
&&\chi^{2} (\{\zeta_k\})\nonumber\\&& = 
\displaystyle\sum_{i} 
\frac{\left[ \mathcal{E}_{i} - 
\mathcal{T}_{i} \left(\{\zeta\}\right) 
\left(1 - \sum_{j} 
\gamma_{j}^{i} {b}_{j}\right) 
\right]^{2}  }
{\delta^{2}_{{i}, \mathrm{unc}} 
\mathcal{T}_{i}^{2}(\{\zeta\})+ 
{\delta}^{2}_{i, \mathrm{stat}} 
\mathcal{E}_{i} \mathcal{T}_{i} 
(\{\zeta\}) 
\left(1 - \sum_{j} \gamma^{i}_{j} 
{b}_{k}\right)} \nonumber  \\ 
&& + \sum_{i} \ln \frac{\delta^{2}_{{i}, 
\mathrm{unc}} 
\mathcal{T}_{i}^{2} (\{\zeta\}) + 
\delta^{2}_{{i}, \mathrm{stat}} 
\mathcal{E}_{i} \mathcal{T}_{i}  
(\{\zeta\}) } 
{ \delta^{2}_{{i}, \mathrm{unc}} 
\mathcal{E}_{i}^{2} + 
\delta^{2}_{{i}, \mathrm{stat}} 
\mathcal{E}^{2}_{i}}  + 
\sum_{j} b_{j}^{2}\,,
\end{eqnarray}
with $\mathcal{E}$ is  the measured experimental value, 
and $\mathcal{T}$ is the 
theoretical prediction based on the fit parameters $\{\mathbf{\zeta}_k\}$. 
The parameters ${\delta}_{i,\mathrm{stat}}$  , $\delta_{i,\mathrm{unc}}$, 
and $\gamma^{i}_{j}$ denote the relative statistical, 
uncorrelated systematic, 
and correlated systematic uncertainties, respectively. 
The nuisance parameters ${b}_{{k}}$ are related to 
correlated systematic uncertainty and are 
determined with the $\{\zeta_k\}$ 
parameters simultaneously in the QCD fit.

The above ${\chi}^{2}$ function is incorporated in the {\tt xFitter} framework, 
in conjunction with other tools required for a perturbative 
QCD analysis of diffractive PDFs. 
One then can use this package to perform all the essential operations 
such as DGLAP evolution up to NNLO accuracy, 
theoretical calculation of the relevant observables and 
finding out optimal parameter values 
and deducing their uncertainty collectively. 
As we specified above in order to find the optimal fit 
parameter values one needs 
to minimize the ${\chi}^{2}$ function, 
this is achieved by utilizing the {\tt MINUIT} CERN package~\cite{James:1975dr}. 
This package finds 
the parameter uncertainties by considering the $\chi^{2}$ function around its minimum. 
{\tt MINUIT} has five minimization algorithms, here we choose to 
work with \texttt{MIGRAD} which is the 
most commonly used method of minimization. 

In practice, we need to propagate these parameter 
uncertainties to the observables or other 
quantities like the diffractive PDFs themselves. 
For this aim, a set of eigenvector PDF sets along with the  
the central values of the diffractive PDFs is formed,
for our fit that has 9 free parameters, 
the total number of PDF sets will be 19. 
Each member of the error set is derived by either increasing or 
decreasing one of the parameters by its uncertainty. 
Then, the uncertainty of a quantity ${\cal O}$ due to its dependence on 
PDFs is given as in Ref.~\cite{Nadolsky:2008zw},
\begin{eqnarray}
\Delta {\cal O} = 
\frac{1}{2} 
\sqrt{\sum_{{i}={1}}^{N} 
\left({\cal O}_{i}^{(+)} - 
{\cal O}_{i}^{(-)}
\right)^{2}}\,,
\end{eqnarray}
where ${\cal O}_{i}^{(\pm)}$ refer to 
the values of ${\cal O}$ which are 
calculated from PDF sets of the $i$-th parameter 
along with the $\pm$ directions. 
In the derivation of this relation, 
it is assumed that the variation of ${\cal O}$ can be 
approximated by linear terms of its 
Taylor series and then the gradient is approximated, 
which produces the above result. 
In the next section, we present our main results and findings of 
the {\tt SKMHS22}  diffractive PDFs and 
some observables in order to show the quality of the analysis.

%
\section{Fit results}\label{sec:results}

This section includes the main results of the {\tt SKMHS22} 
diffractive PDFs analysis. 
As we discussed in detail earlier, in this QCD analysis we 
present three different QCD fits to 
determine the diffractive PDFs which are {\tt SKMHS22-tw2}, 
{\tt SKMHS22-tw2-tw4} and {\tt SKMHS22-tw2-tw4-RC}.
In this section we will present and discuss these results in turn. 
The similarity and difference of these results 
over different kinematical ranges
will be highlighted, and the stability of the results upon the 
inclusion of 
higher-twist corrections will be discussed in detail. 
This section also includes a detailed comparison 
with the diffractive DIS data analyzed in this work.

The best fit parameters for three sets of {\tt SKMHS22} diffractive PDFs are 
shown in Table~\ref{tab-dpdf-all} along with their experimental errors.
Considering the numbers presented in this table, 
some comments are in order.
The parameters 	$\eta_g$ and 	$\eta_q$   are 
considered to be fixed at zero as the present
diffractive DIS data sets do not have enough power 
to constrain all shape parameters of the 
distributions.

The parameter \{$w_i$\} for the Reggeon structure function 
for the {\tt SKMHS22-tw2-tw4-RC} analysis 
presented in  Eq.~\ref{eq:Reggeonstructure} 
are determined along the fit 
parameters and then keep fixed to their best fitted values. 
The parameters $w_4$ and $w_5$ are keep fixed at zero. 
The values for 
the strong coupling constant $\alpha_s(M_Z^2)$ and the 
charm and bottom quark 
masses also are shown in the table as well.  

%
%
\begin{table*}[ht]
\begin{center}
\caption{\small Best fit parameters obtained with the {\tt SKMHS22-tw2}, 
{\tt SKMHS22-tw2-tw4} and {\tt SKMHS22-tw2-tw4-RC} fits at the initial 
scale of $Q_{0}^{2} = 1.8 \, {\text{GeV}}^2$ along with their 
experimental uncertainties. The values marked with the ({*}) are fixed in the fit. }
\begin{tabular}{ c | c | c  | c  }
\hline \hline
			Parameters	  & {\tt SKMHS22-tw2} & {\tt SKMHS22-tw2-tw4}    &    {\tt SKMHS22-tw2-tw4-RC}  \\  \hline \hline
			$\alpha_g$    & $1.00 \pm 0.16$ & $1.07 \pm 0.17$  & $1.43 \pm 0.23$  \\ 
			$\beta_g$     & $0.226 \pm 0.066$ & $0.332 \pm 0.070$  & $0.447 \pm 0.070$ \\ 
			$\gamma_g$    & $0.27 \pm 0.15$   & $0.19 \pm 0.14$   & $0.37 \pm 0.14$\\ 
			$\eta_g$      & $0.0^*$           & $0.0^*$            & $0.0^*$\\   
			$\alpha_q$    & $0.305 \pm 0.022$ & $0.517 \pm 0.041$  & $0.727 \pm 0.059$ \\ 
			$\beta_q$     & $1.474 \pm 0.069$ & $1.887 \pm 0.081$  & $2.149 \pm 0.0584$\\ 
			$\gamma_q$    & $0.509 \pm 0.034$ & $0.980 \pm 0.0948$  & $1.137 \pm 0.050$\\ 
			$\eta_q$      & $0.0^*$           & $0.0^*$            & $0.0^*$ \\     
			$\alpha_{\pom}(0)$  & $1.0934 \pm 0.0032$  & $1.1021 \pm 0.0037$ & $1.0965 \pm 0.0037$  \\
			$\alpha_{\reg}(0)$  & $0.316 \pm 0.053$    & $0.400 \pm 0.053$   & $0.418 \pm 0.054$  \\
			$A_{\reg}$          & $21.7 \pm 5.7$       & $15.0 \pm 3.9$       &  $13.2 \pm 3.5$ \\   
			$w_1$          & $0.0^*$       & $0.0^*$       & $0.23^*$  \\   
			$w_2$          & $0.0^*$       & $0.0^*$       & $3.79^*$   \\   
			$w_3$          & $0.0^*$       & $0.0^*$       & $14.9^*$  \\   
			$w_4$          & $0.0^*$       & $0.0^*$       & $0.0^*$   \\  
			$w_5$          & $0.0^*$       & $0.0^*$       & $0.0^*$  \\ \hline 
			$\alpha_s(M_Z^2)$   & $0.1185^*$           & $0.1185^*$          &  $0.1185^*$  \\
			$m_c$               & $1.40^*$             & $1.40^*$            &  $1.40^*$  \\
			$m_b$               & $4.75^*$             & $4.75^*$            &  $4.75^*$  \\ 	\hline \hline
		\end{tabular}
		\label{tab-dpdf-all}
	\end{center}
\end{table*}
%
%

\begin{figure}[htb]
\vspace{0.50cm}
\resizebox{0.480\textwidth}{!}{\includegraphics{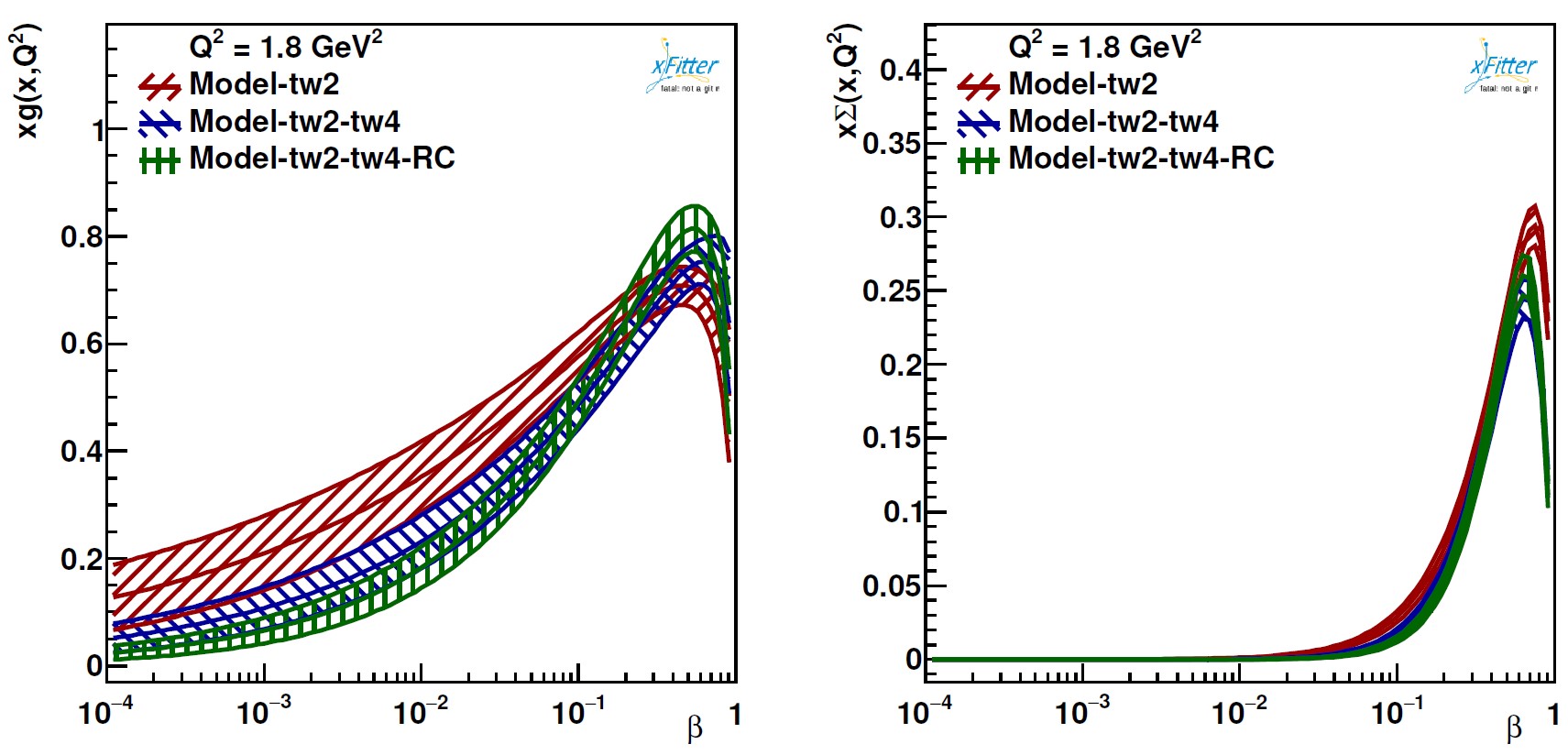}} 	
\resizebox{0.480\textwidth}{!}{\includegraphics{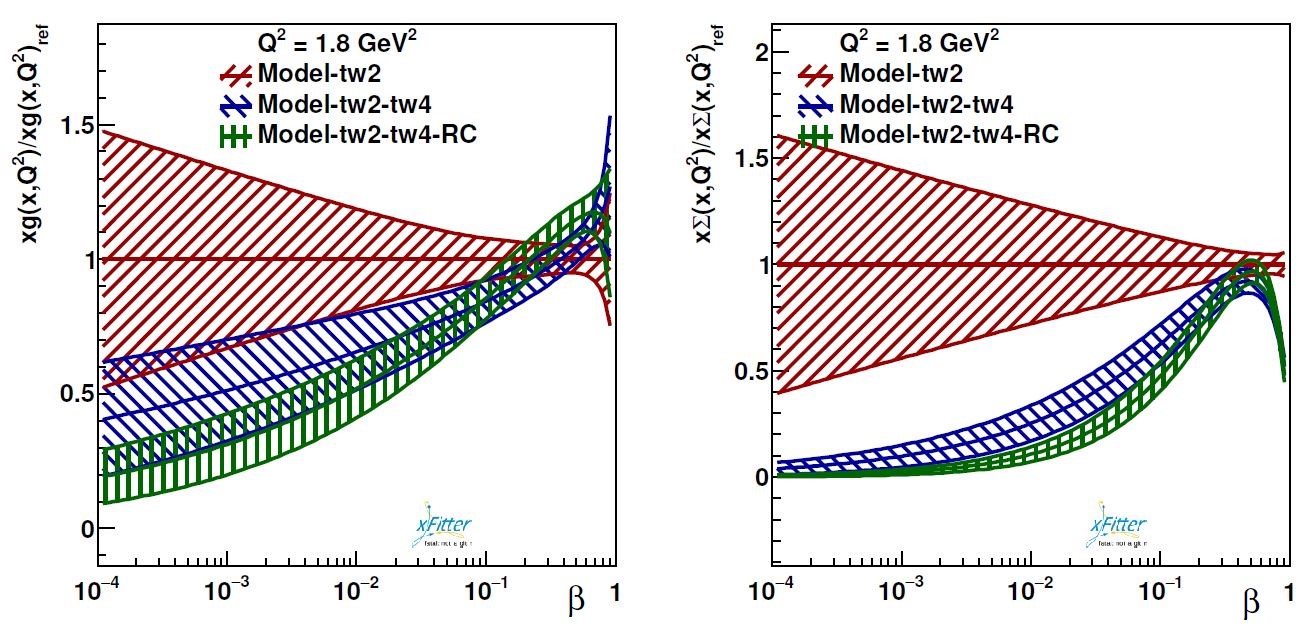}} 	
\begin{center}
\caption{ \small 
The {\tt SKMHS22} diffractive PDFs at the input scale of $Q_0^2 = 1.8$ GeV$^2$. 
All three diffractive PDFs sets are shown for comparison.  
The lower panels represent the ratio of {\tt SKMHS22-tw2-tw4} and 
{\tt SKMHS22-tw2-tw4-RC} to the {\tt SKMHS22-tw2}.}
\label{fig:DPDF-Q0}
\end{center}
\end{figure}

\begin{figure}[htb]
	\vspace{0.50cm}
	\resizebox{0.480\textwidth}{!}{\includegraphics{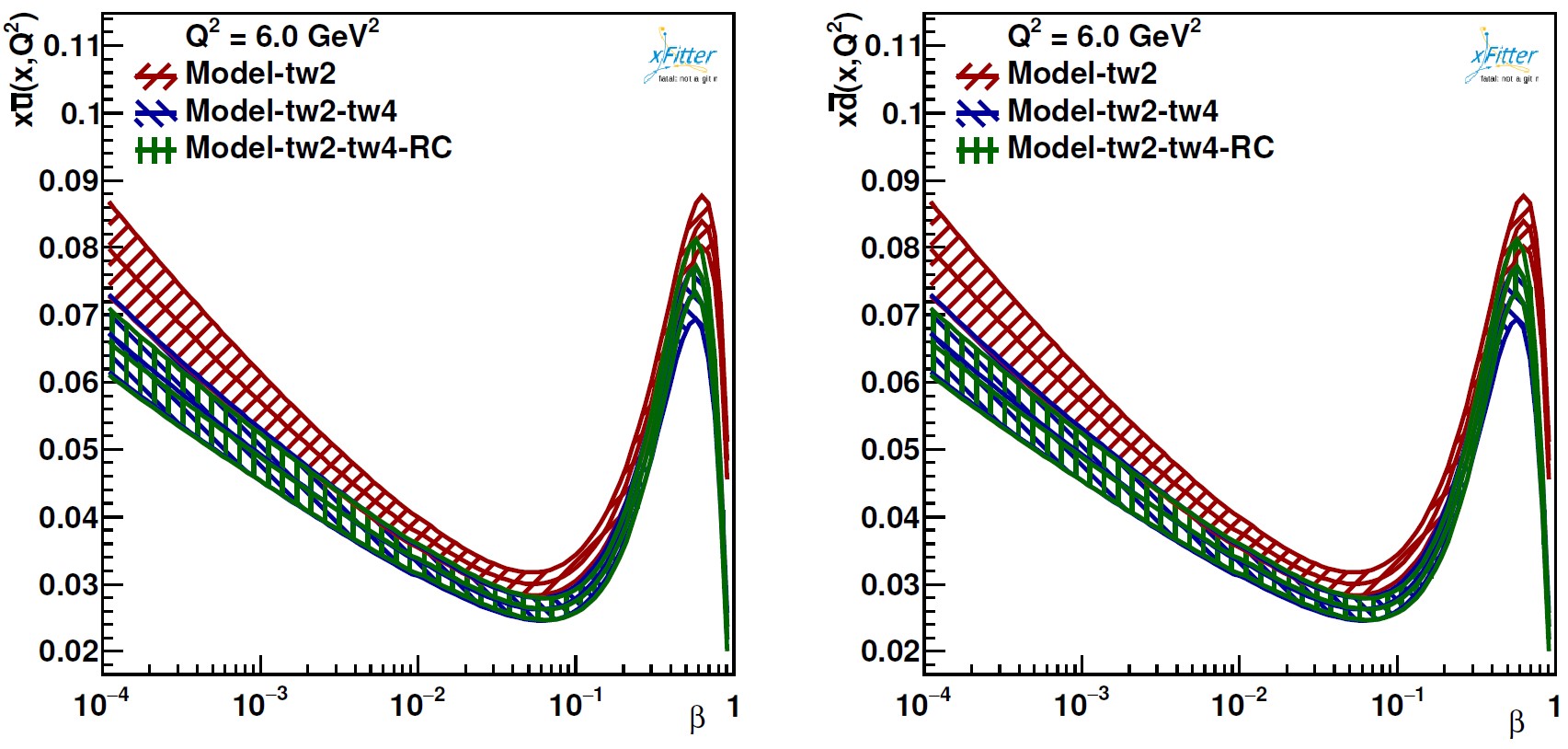}} 	
	\resizebox{0.480\textwidth}{!}{\includegraphics{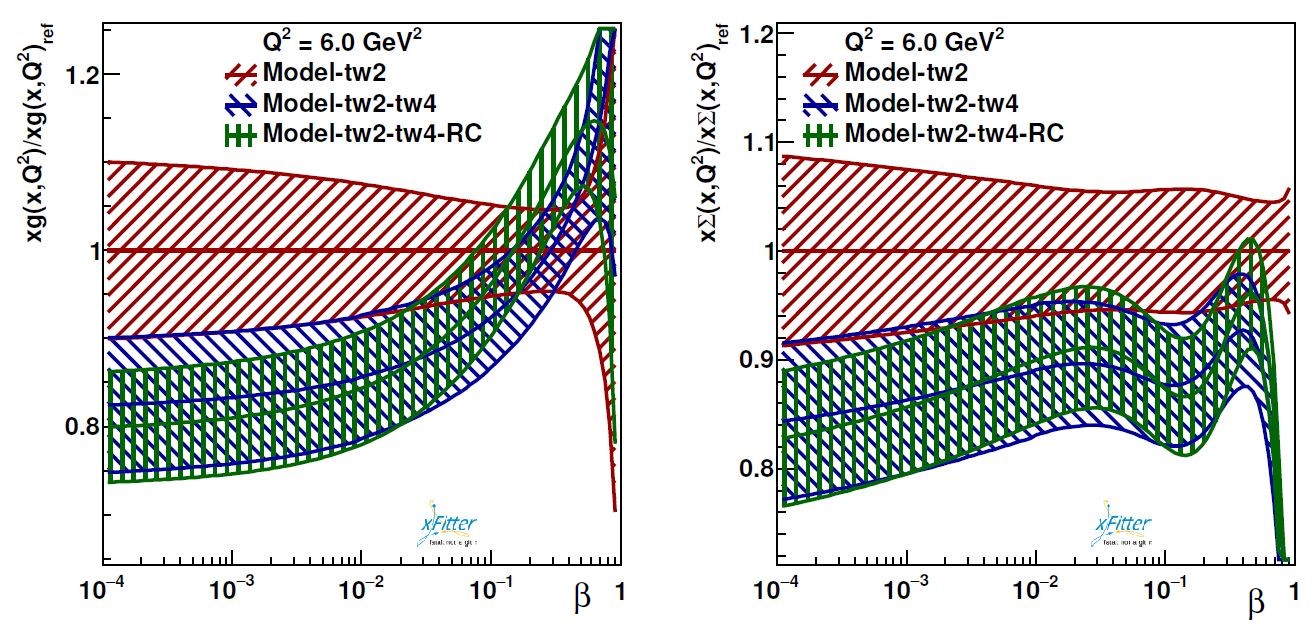}} 	
	\begin{center}
		\caption{ \small 
Same as in Fig.~\ref{fig:DPDF-Q0} but this time for the the higher energy scale of 6 GeV$^2$.
		}
		\label{fig:DPDF-Q6}
	\end{center}
\end{figure}

\begin{figure}[htb]
	\vspace{0.50cm}
	\resizebox{0.480\textwidth}{!}{\includegraphics{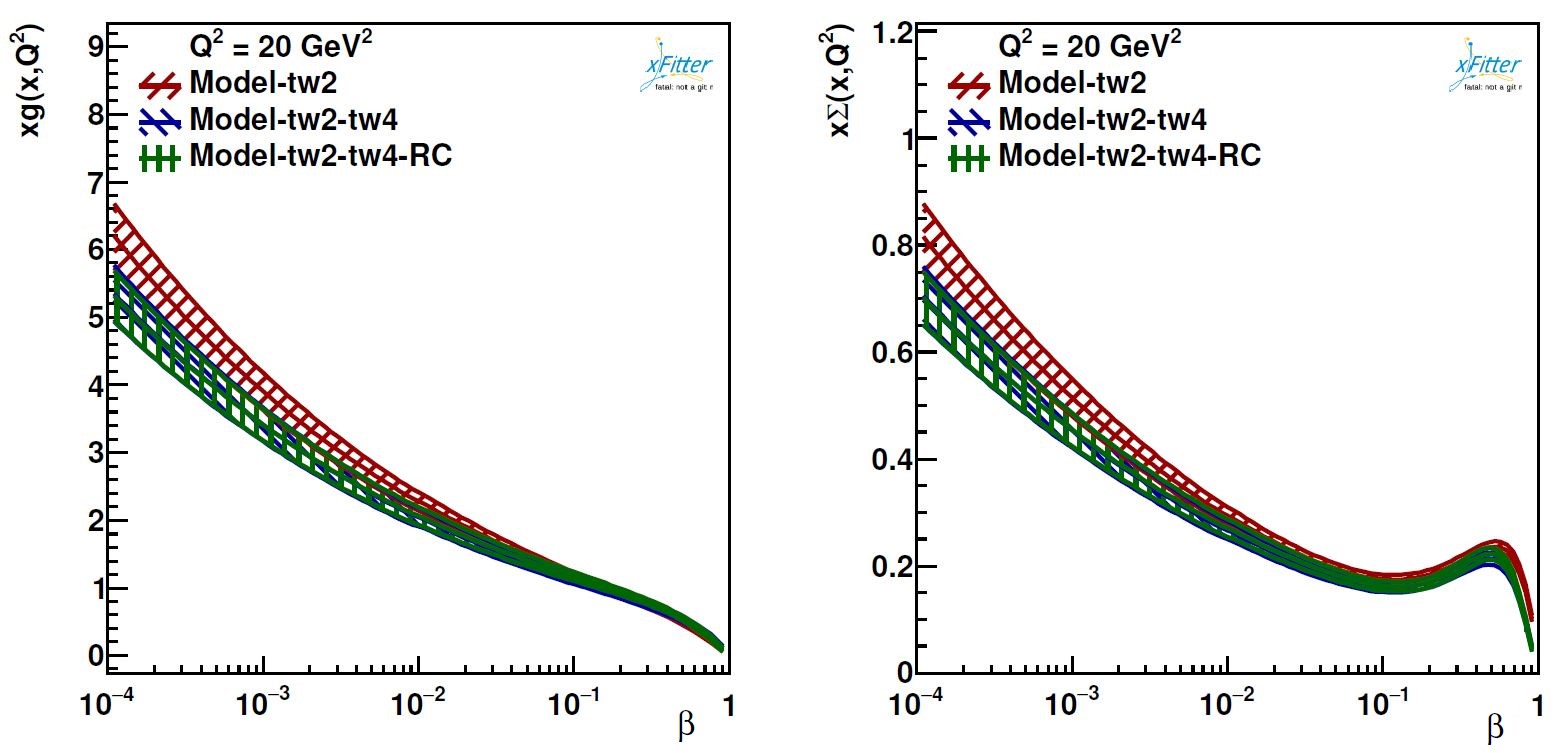}} 	
	\resizebox{0.480\textwidth}{!}{\includegraphics{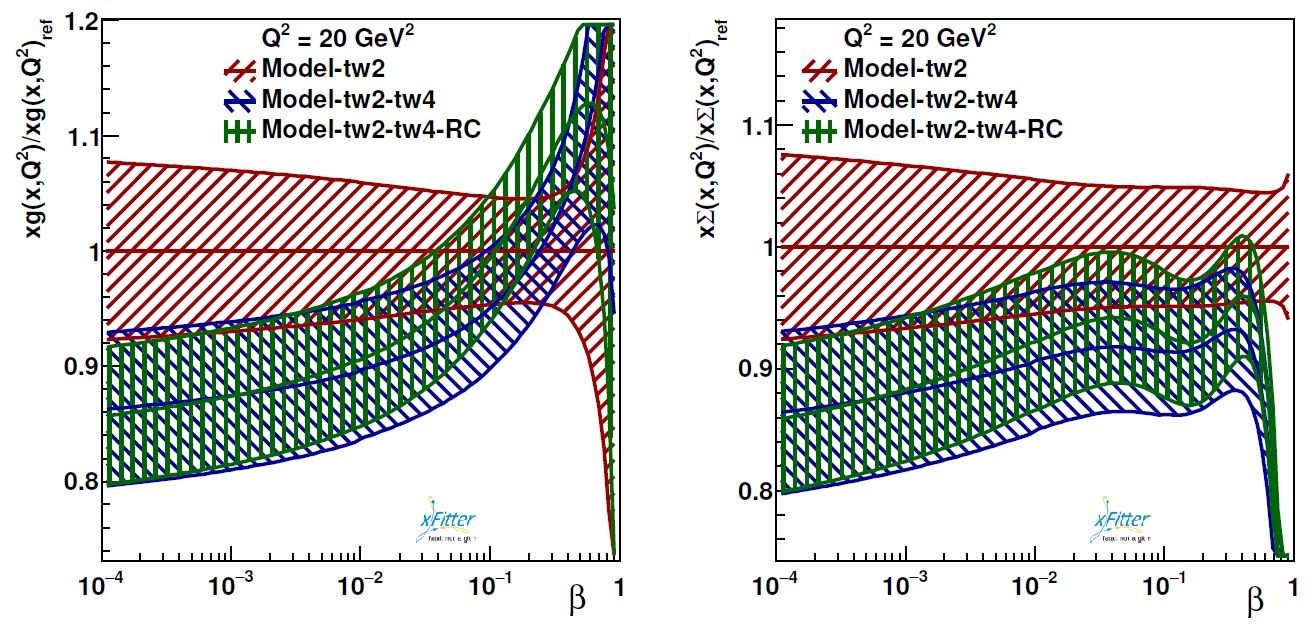}} 	
	\begin{center}
		\caption{ \small 
Same as in Fig.~\ref{fig:DPDF-Q0} but this time for the the higher energy scale of 20 GeV$^2$.
		}
		\label{fig:DPDF-Q20}
	\end{center}
\end{figure}

In the following, we turn our attention to the detailed comparison 
of the three {\tt SKMHS22} diffractive PDFs sets.
We display the {\tt SKMHS22-tw2}, {\tt SKMHS22-tw2-tw4} and {\tt SKMHS22-tw2-tw4-RC} 
diffractive PDFs parameterized in our QCD fits, Eqs.~\eqref{eq:parametrizationformq}  and 
\eqref{eq:parametrizationformg}, along with their uncertainty bands in 
Fig.~\ref{fig:DPDF-Q0} at the input scale $Q_0^2 = 1.8$ GeV$^2$. 
The higher Q$^2$ value results at 6 and 20~GeV$^2$ are shown in Figs.~\ref{fig:DPDF-Q6} and 
Fig.~\ref{fig:DPDF-Q20}, respectively.
The lower panels show the ratio of 
{\tt SKMHS22-tw2-tw4} and {\tt SKMHS22-tw2-tw4-RC} to the {\tt SKMHS22-tw2}.

\begin{figure}[htb]
\vspace{0.50cm}
\resizebox{0.480\textwidth}{!}{\includegraphics{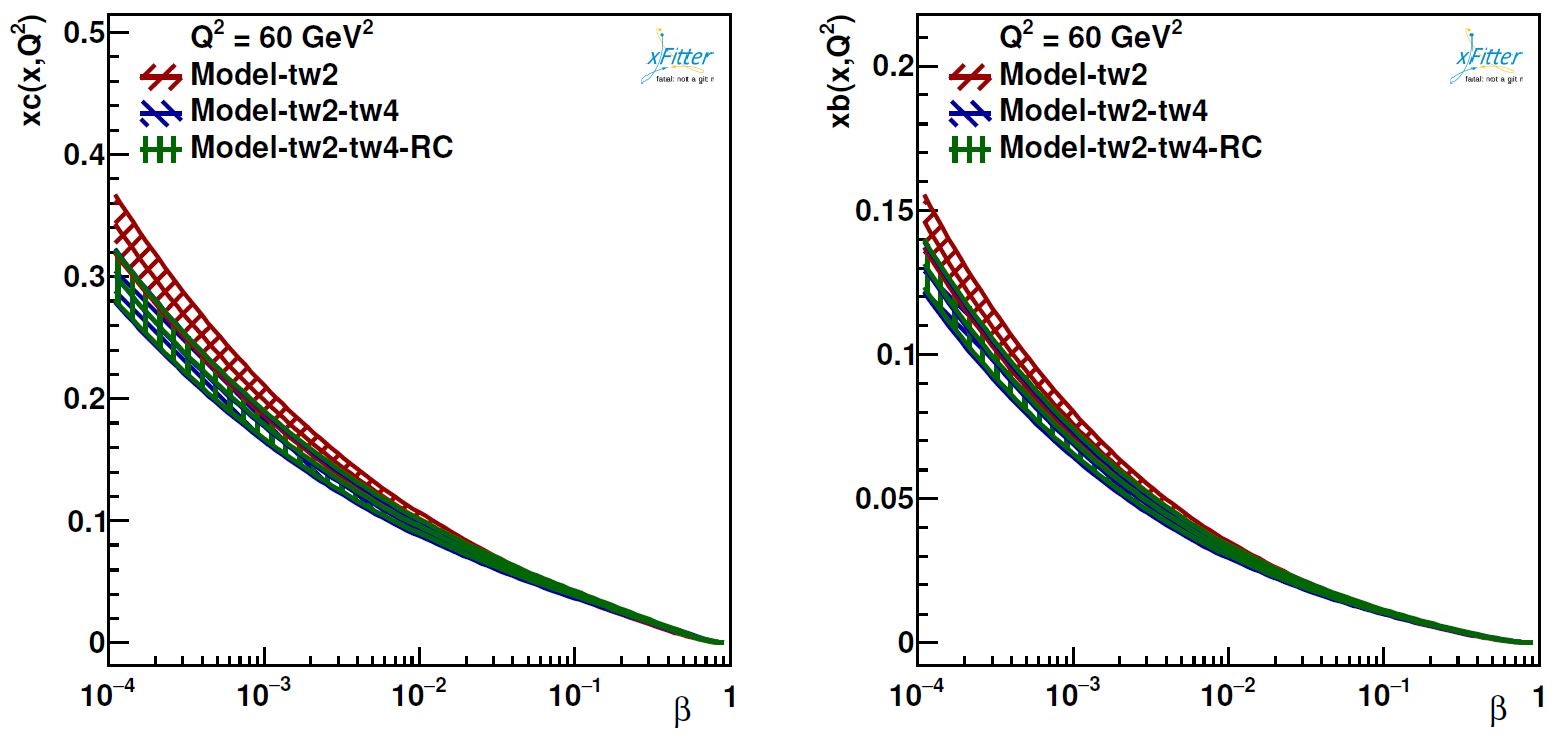}} 	
\resizebox{0.480\textwidth}{!}{\includegraphics{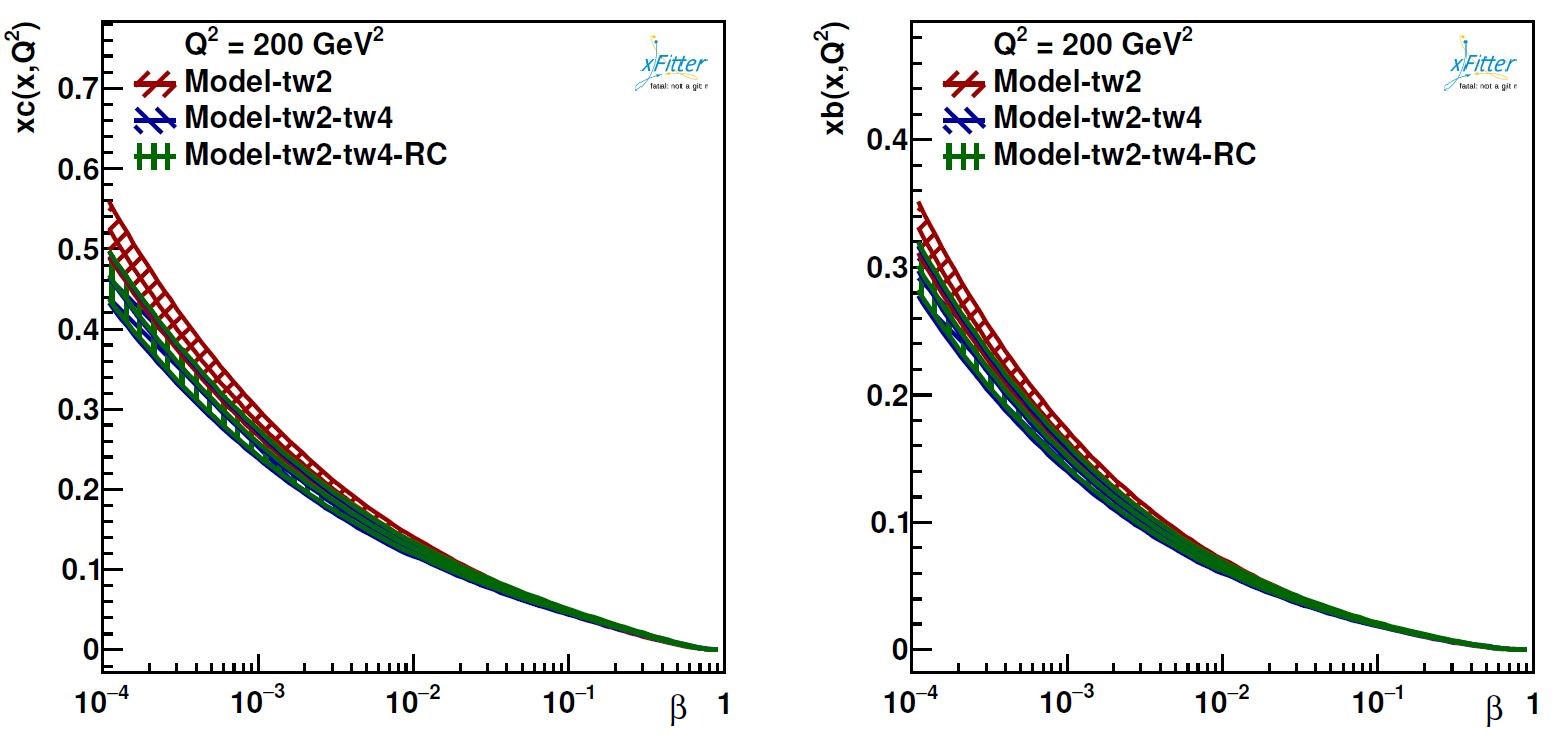}}
\begin{center}
\caption{ \small 
The perturbatively generated {\tt SKMHS22} diffractive PDFs for the 
charm and bottom quark densities along with their error bands at the scale of $Q^2 = 60$ 
and 200~GeV$^2$. All three diffractive PDFs sets are shown for comparison. 
}
\label{fig:DPDF-c-b}
\end{center}
\end{figure}

In Fig.~\ref{fig:DPDF-c-b}, we display the perturbatively generated {\tt SKMHS22} 
diffractive PDFs for the charm and bottom quark densities along 
with their error bands at the scale of $Q^2 = 60$~GeV$^2$ and 200 GeV$^2$. 
All three diffractive PDFs sets are shown for comparison. 

A remarkable feature of the {\tt SKMHS22} diffractive gluon and quark PDFs shown 
in these figures is the difference both in the shape and error bands, 
which reflects the effect arising from the inclusion of 
higher twist corrections.
As can be seen, for all cases the inclusion of the
twist-4 and  Reggeon corrections lead to 
slightly small error bands. 
This is consistent with the ${\chi^{2}}$ value presented in Table~\ref{tab:chi2-all}.
The inclusion of the twist-4 corrections and Reggeon contributions leads to 
an enhancement of the gluon diffractive PDFs at high $\beta$, 
and a reduction of the singlet PDFs as well.
Considering such corrections also affects the small 
regions of $\beta$ and leads to the reduction of the central 
values of all distributions. 
These findings indicate that for the given kinematical range of 
the diffractive DIS data from the ZEUS and H1 Collaborations, 
considering the higher twist corrections is of crucial importance.

For the gluon PDFs, we see a very large error band for a 
small region of $\beta$, namely $\beta < 0.01$, which indicates 
that the available diffractive DIS data do not have 
enough power to constrain the low-$\beta$ gluon density. 
To better constrain the gluon PDFs, the diffractive dijet productions data 
need to be taken into account~\cite{H1:2014pjf,H1:2015okx}. 
In terms of future work, it would be very  interesting to 
repeat the QCD  analysis described here to study the impact of 
diffractive dijet production data on the diffractive 
gluon PDF and its uncertainty.

The same discussion also holds for the case of 
heavy quark diffractive PDFs. 
As one can see from Fig.~\ref{fig:DPDF-c-b}, the 
{\tt SKMHS22-tw2-tw4-RC} charm and bottom quark densities 
lie below other results for all regions of $\beta$.

In Table.~\ref{tab:chi2-all}, 
we present the values for the $\chi^2$ per 
data point for both the individual and the 
total diffractive DIS data sets included in the {\tt SKMHS22} analysis.
The values are shown for all three sets of {\tt SKMHS22-tw2}, 
{\tt SKMHS22-tw2-tw4} and {\tt SKMHS22-tw2-tw4-RC}.
Concerning the fit quality of the total diffractive DIS data sets, 
the most noticeable feature is an improvement in 
the inclusion of the Reggeon contribution. 
The improvement of the individual  $\chi^2$ per data point is particularly 
pronounced for the H1-LRG 2012 when the Reggeon contribution in 
considered in the analysis. 
This finding demonstrates the fact that the  inclusion
of the higher-twist corrections
improves the description of
the diffractive DIS data sets.

%
\begin{table*}[htb]
	\caption{ \small The values of $\chi^2/N_{\text{pts}}$ for the data sets included in the {\tt SKMHS22} global fits. } \label{tab:chi2-all}
	\begin{tabular}{l c c c  c }
		\hline \hline
		& ~~{\tt SKMHS22-tw2}~~            & ~~{\tt SKMHS22-tw2-tw4}~~  &  ~~{\tt SKMHS22-tw2-tw4-RC}~~  \\ \hline
		\text{Experiment} & ${\chi}^2/{N}_{\text{pts}}$  & ${\chi}^{2}/{N}_{\text{pts}}$  & ${\chi}^2/{N}_{\text{pts}}$ 
		\tabularnewline
		\hline
		\text{H1-LRG-11} $\sqrt{s} = {225}$~GeV~\cite{H1:2011jpo} & 11/13 &  11/13  & 12/13    \\	
		\text{H1-LRG-11} $\sqrt{s} = {252}$~GeV~\cite{H1:2011jpo} & 20/12 &  21/12  &  19/12\\		
		\text{H1-LRG-11} $\sqrt{s} = {319}$~GeV~\cite{H1:2011jpo} & 6.6/12 &  3.7/12 & 4.6/12\\			
		\text{H1-LRG-12}~\cite{H1:2012pbl}        & 136/165 &  143/165  & 124/165 \\	
		\text{H1/ZEUS combined}~\cite{H1:2012xlc} & 129/100 &  124/100   & 125/100 \\	 \hline
		\text{Correlated} ${\chi}^{2}$  & 11  & 16     & 19 \\  \hline 
		\text{Log penalty} ${\chi}^{2}$  & +11  & +22  & +15 \\  	\hline \hline
		\multicolumn{1}{c}{~\textbf{${\chi}^{2}/{\text{dof}}$}~}  &    $~324/293=1.10~$ &    $~319/293=1.16~$ &   $~319/293=1.08~$ \\  \hline
	\end{tabular}
\end{table*}
%
%

In order further high-light  the {\tt SKMHS22} analysis, 
in the following, we present a detailed comparison of the diffractive DIS data set 
analyzed in this work to the corresponding NLO theoretical predictions 
obtained using the NLO diffractive PDFs from all three sets. 
As we will show, in general, overall good agreements
between the diffractive data and the 
theoretical predictions are achieved for all H1 and ZEUS data.

\begin{figure*}[htb]
\vspace{0.50cm}
\resizebox{0.480\textwidth}{!}{\includegraphics{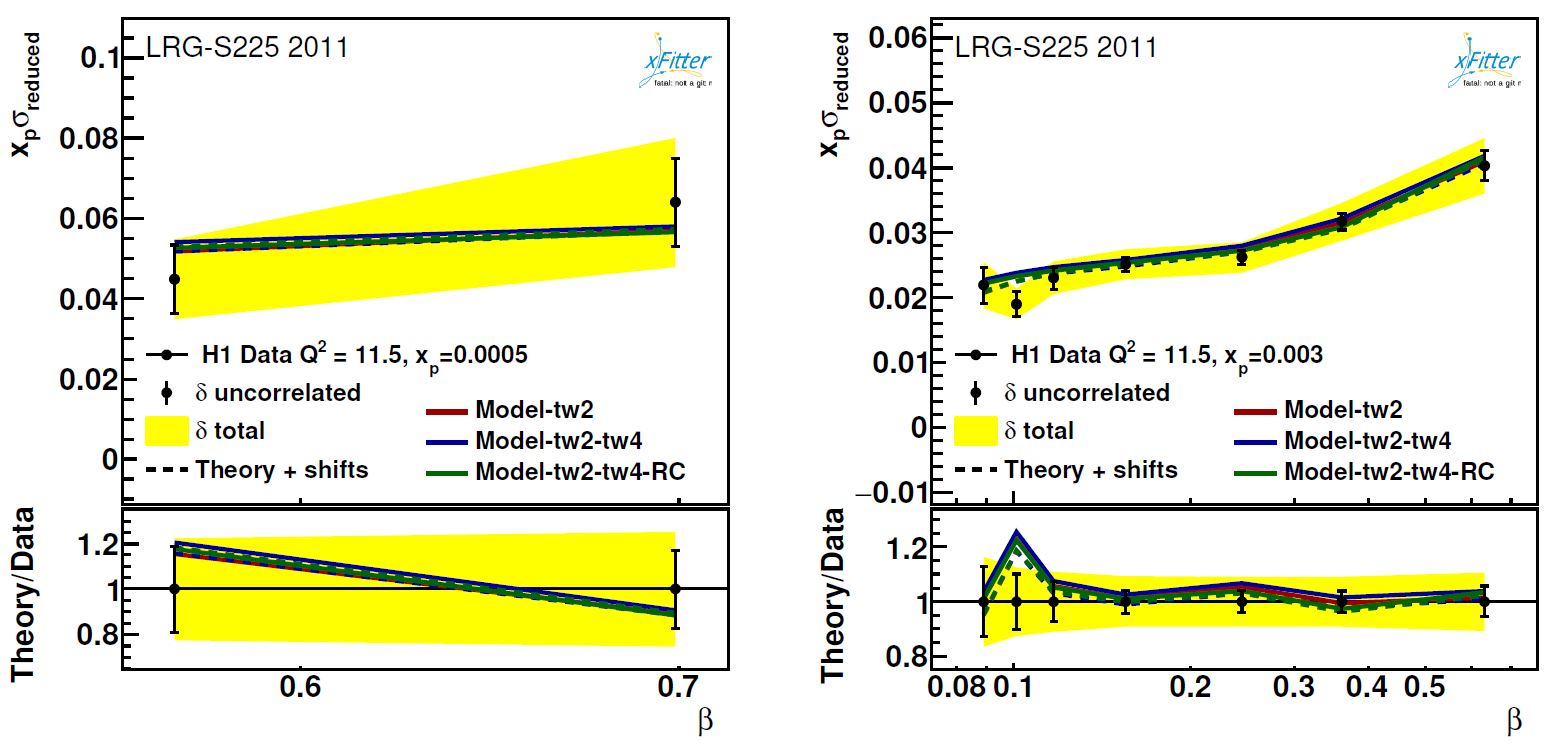}} 	
\resizebox{0.480\textwidth}{!}{\includegraphics{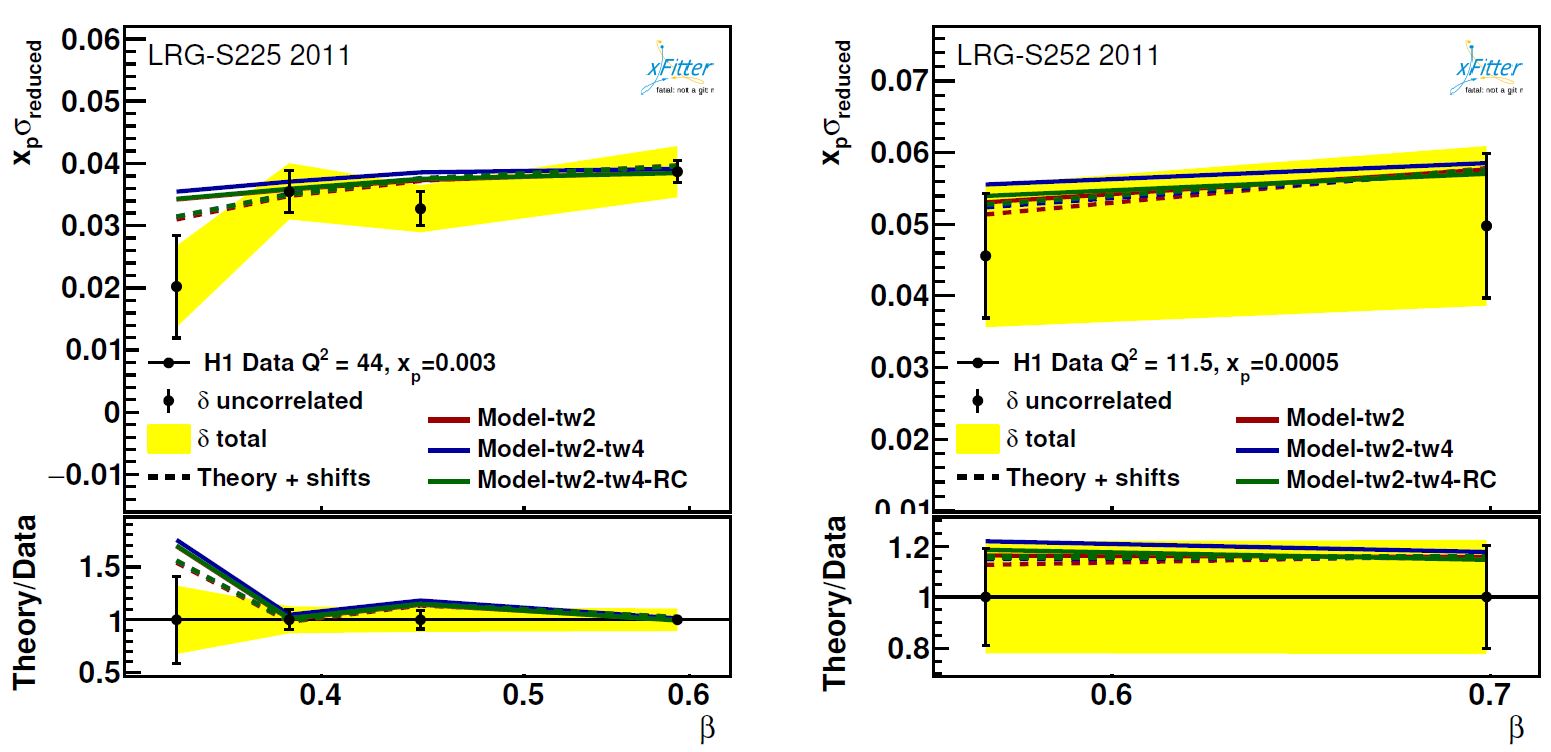}}
\resizebox{0.480\textwidth}{!}{\includegraphics{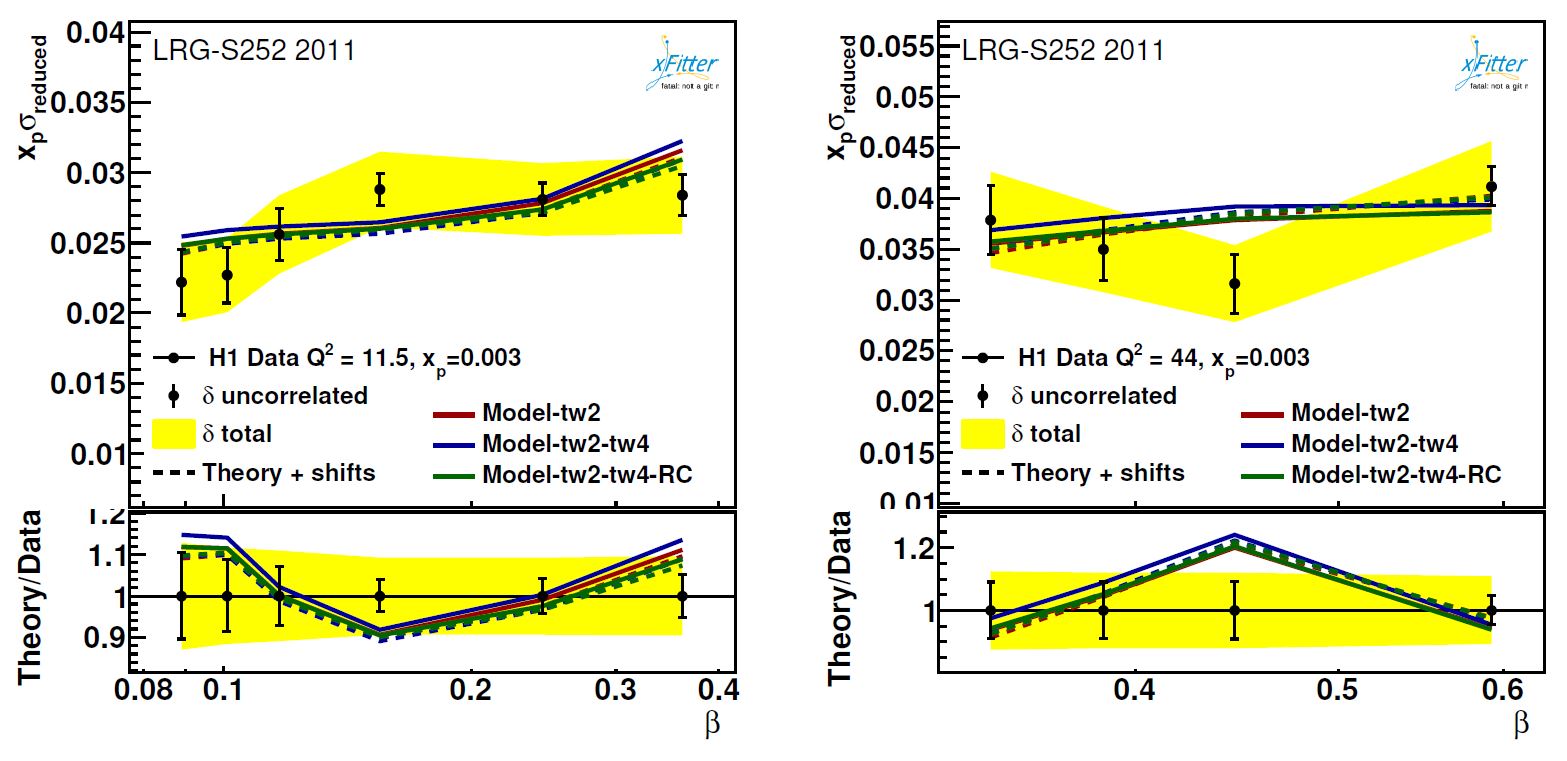}} 	
\resizebox{0.480\textwidth}{!}{\includegraphics{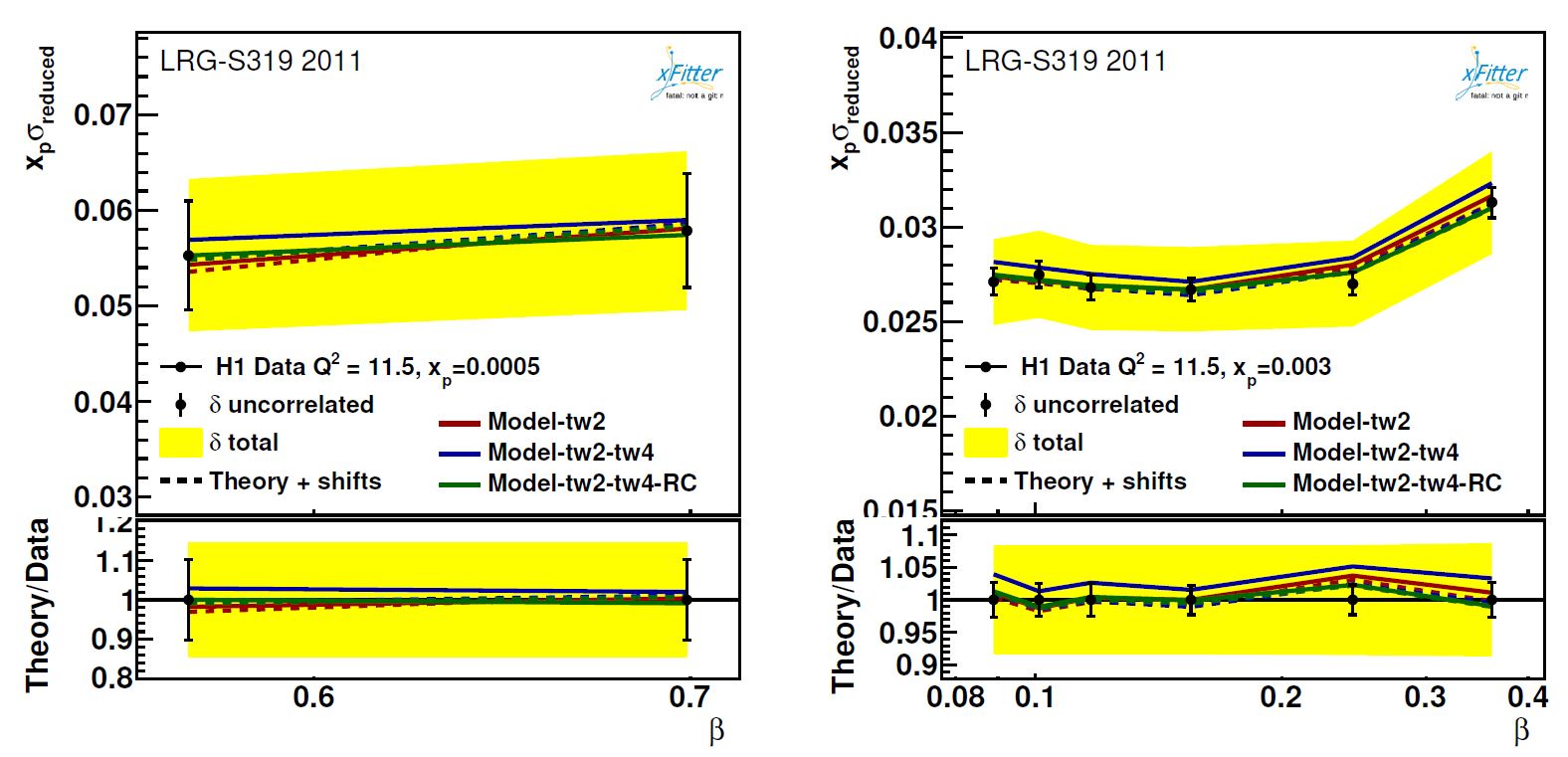}}
\begin{center}
\caption{ \small 
Comparison between the  diffractive DIS data set from the H1-LRG-11 $\sqrt{s} = 225$, 
252 and 329~GeV and the corresponding NLO theoretical predictions 
using all three diffractive PDFs sets. 
We show both the absolute distributions and the data/theory ratios.
}
\label{fig:S225}
\end{center}
\end{figure*}

We start with the comparison to the H1-LRG-11 $\sqrt{s} = 225$, 252 and 319~GeV data. 
In Fig.~\ref{fig:S225}, the {\tt SKMHS22} theory predictions are compared with 
some selected data points. 
The comparisons are shown as a function of $\beta$ and 
for selected bins of $x_{\pom}$=0.0005 and 0.003. 
As one can see, very nice agreement is achieved for all regions of $\beta$

\begin{figure*}[htb]
\vspace{0.50cm}
\resizebox{0.480\textwidth}{!}{\includegraphics{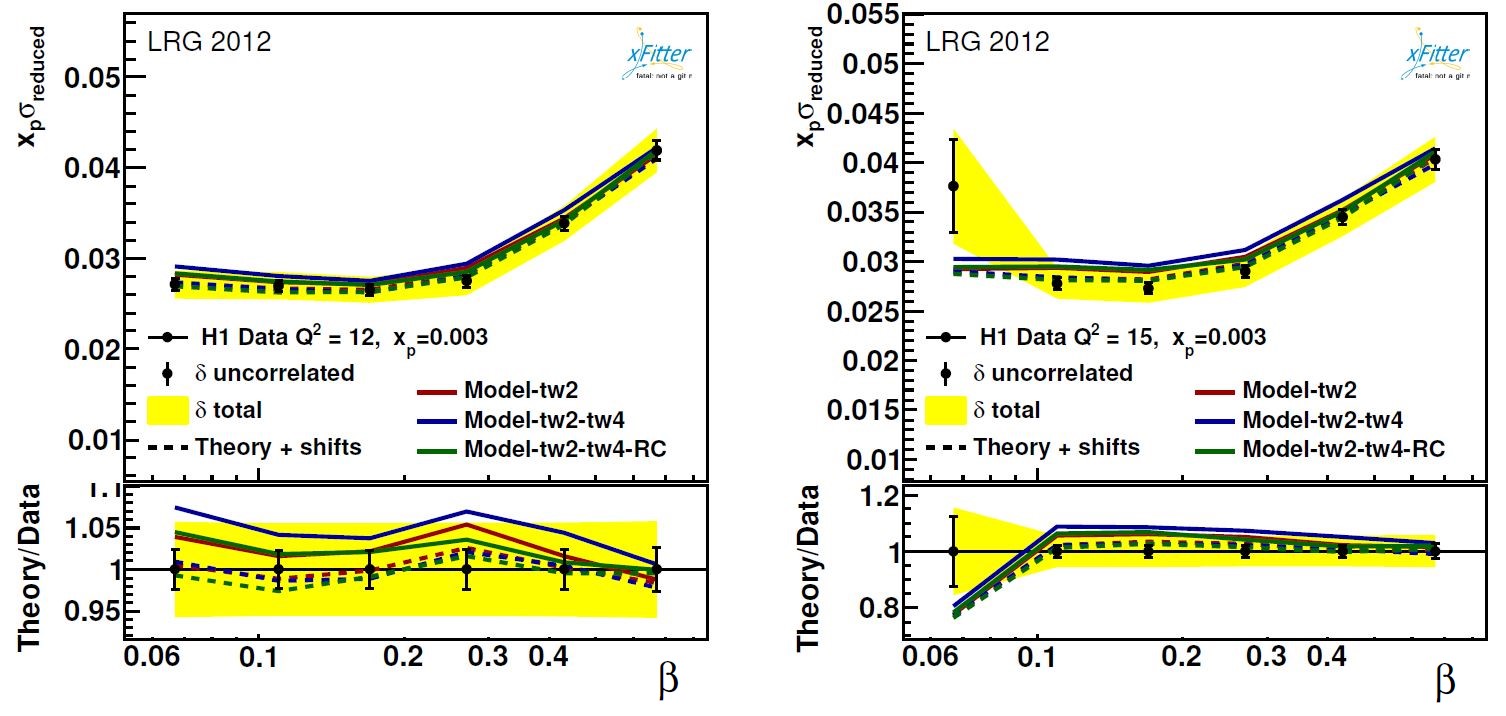}} 	
\resizebox{0.480\textwidth}{!}{\includegraphics{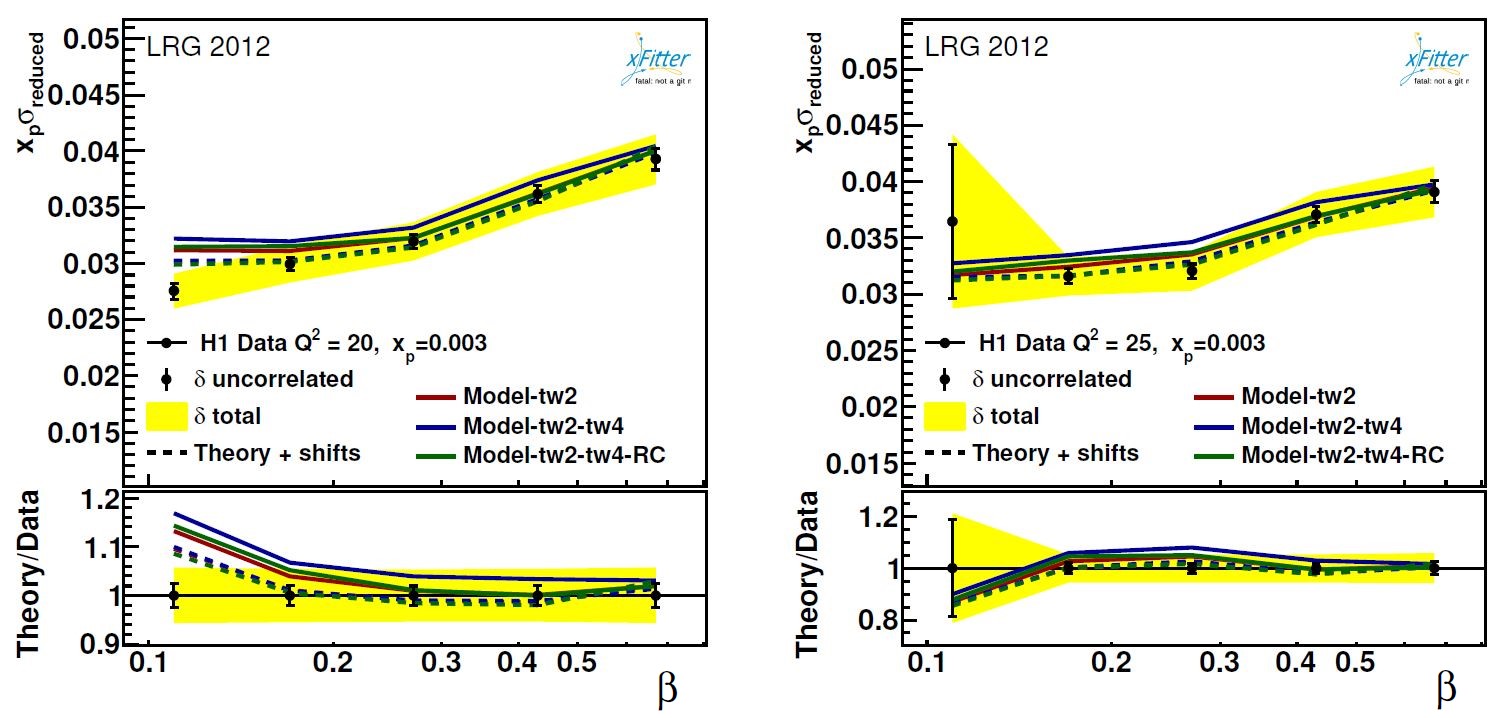}}
\resizebox{0.480\textwidth}{!}{\includegraphics{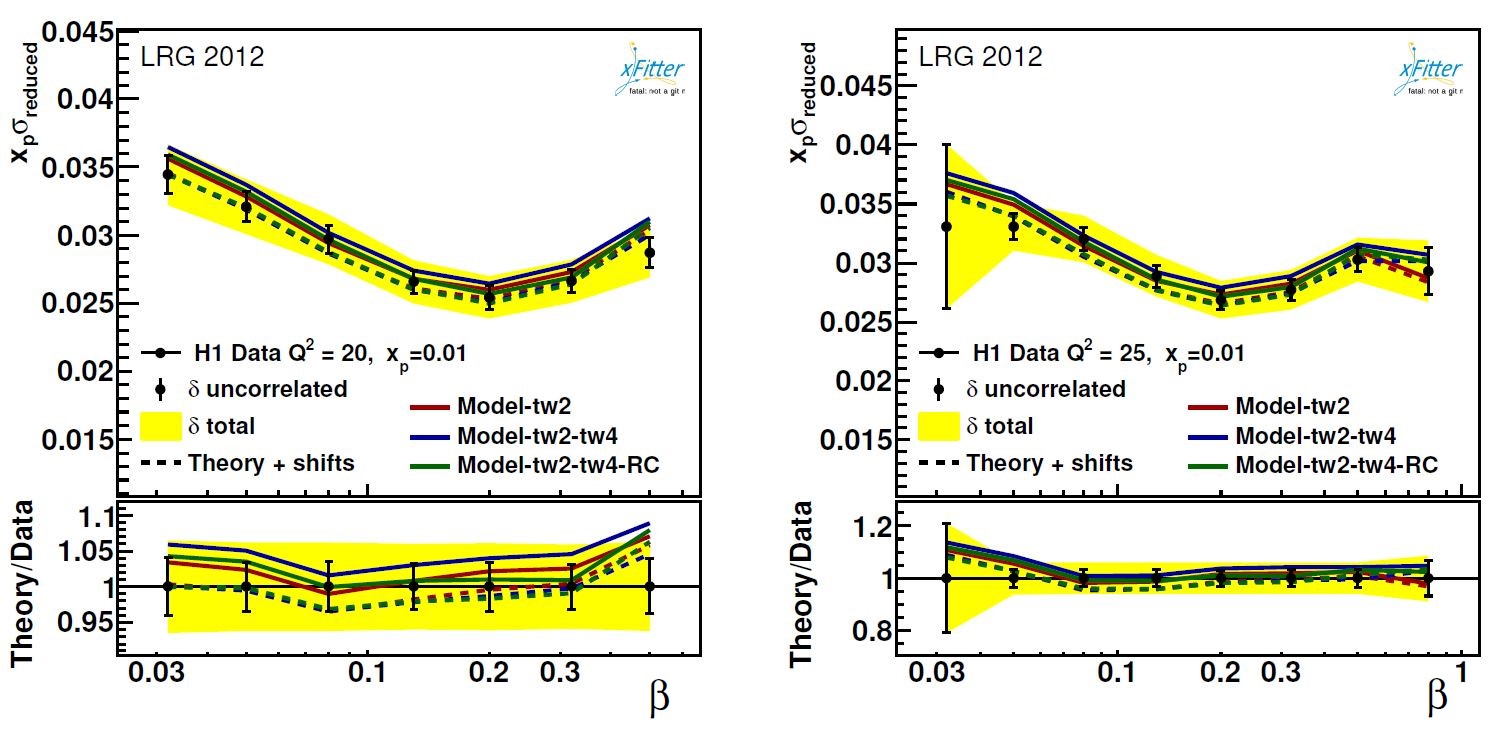}} 	
\resizebox{0.480\textwidth}{!}{\includegraphics{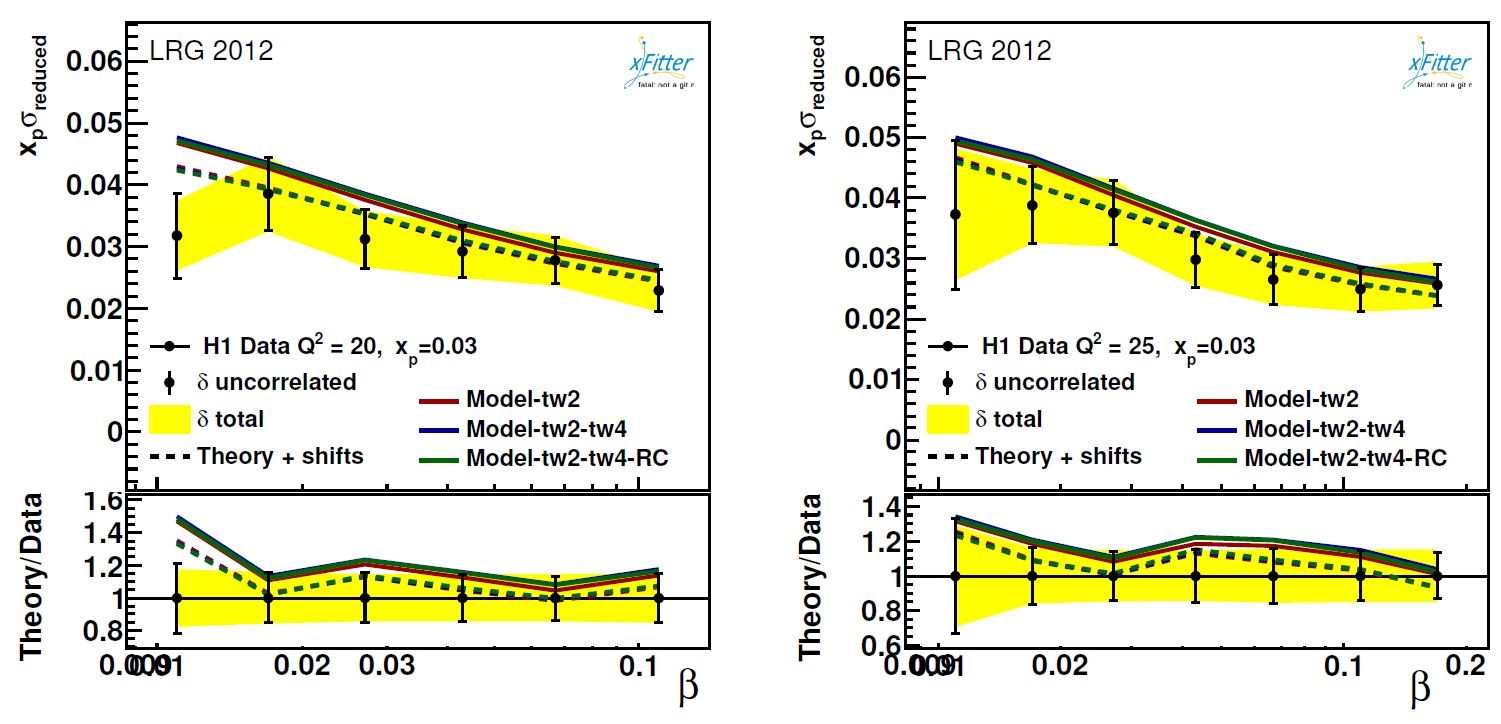}}
\resizebox{0.480\textwidth}{!}{\includegraphics{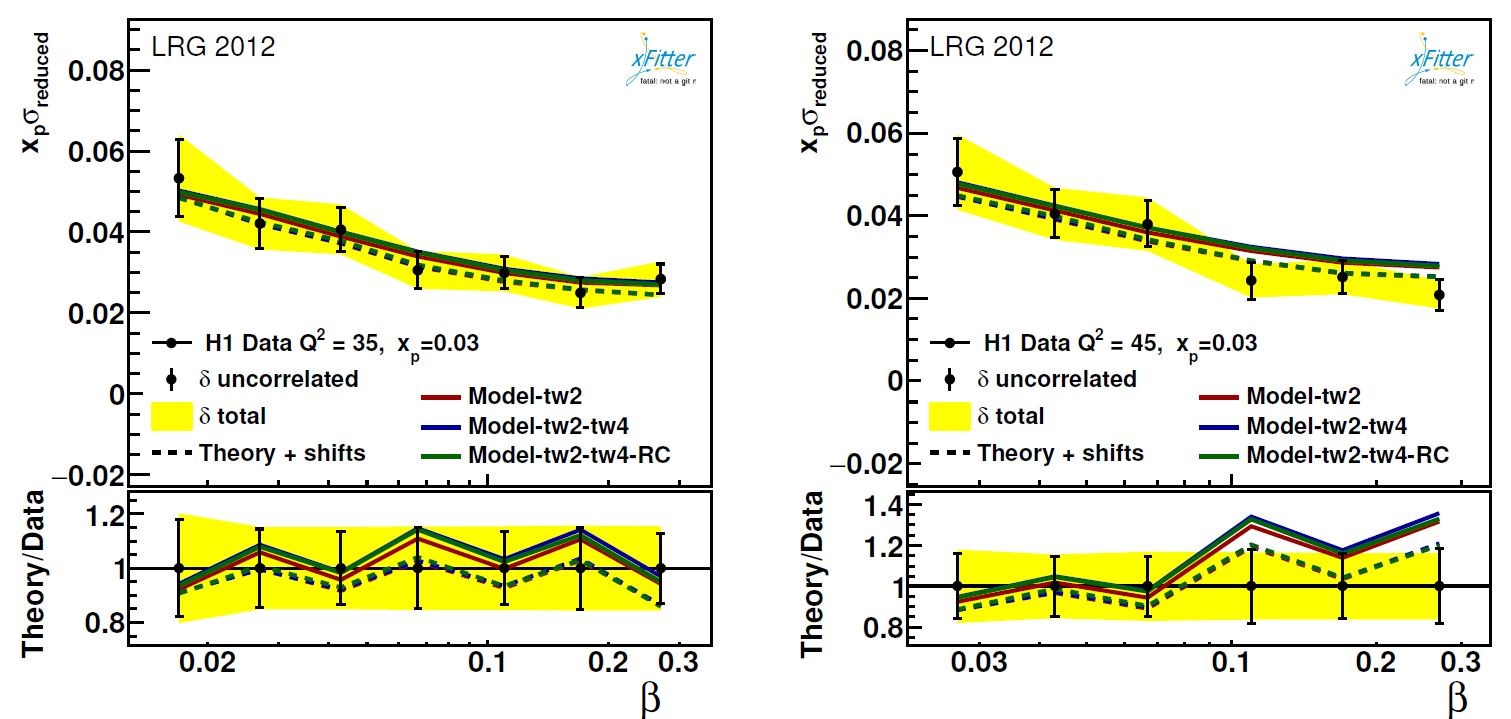}} 	
\resizebox{0.480\textwidth}{!}{\includegraphics{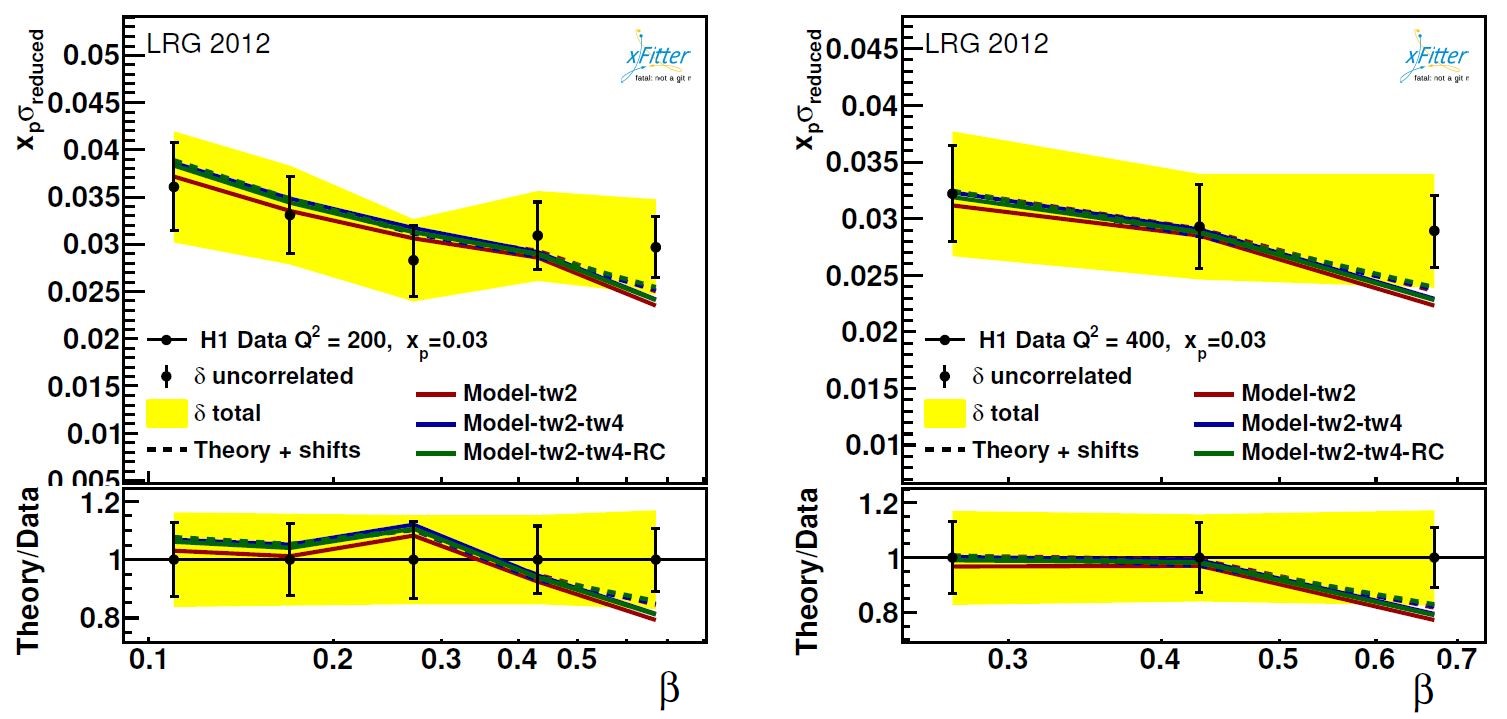}}
\begin{center}
\caption{ \small 
Same as Fig.~\ref{fig:S225} but this time for the  H1-LRG-12 data.
}
\label{fig:H1-LRG-12}
\end{center}
\end{figure*}

The same comparisons are also  shown in Fig.~\ref{fig:H1-LRG-12}  for the H1-LRG-12 for 
some selected bins of $x_{\pom}$ namely 0.01, 0.03, and 0.003.
In these plots, the {\tt SKMHS22} theory precision for 
all three different sets are compared with the H1-LRG-12 diffractive DIS data.
The comparisons are shown as a function of $\beta$ and for
different values of Q$^2$.
As one can see, very good agreements  between the data and 
theory predictions are achieved, consistent with the 
total and individual $\chi^{2}$ presented in Table~\ref{tab:chi2-all}.

\begin{figure*}[htb]
\vspace{0.50cm}
\resizebox{0.480\textwidth}{!}{\includegraphics{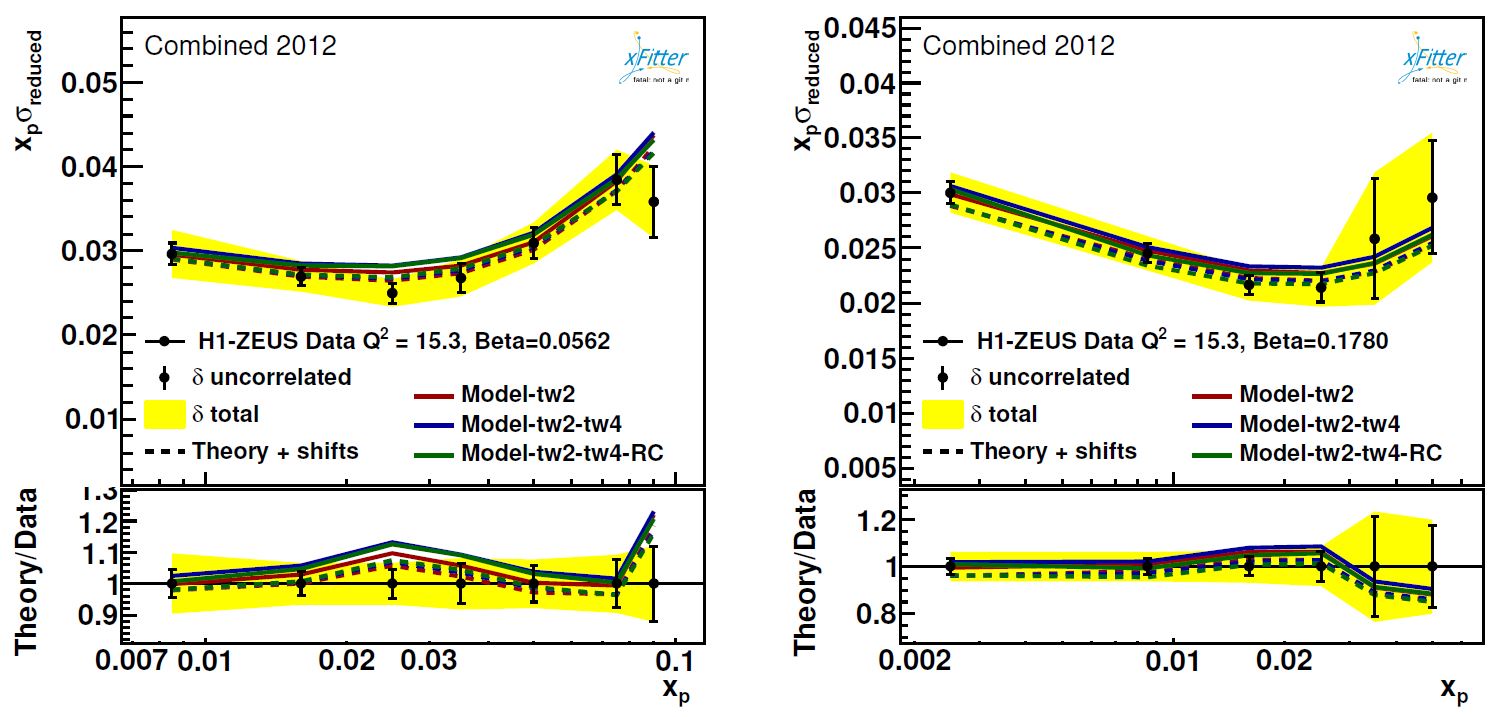}} 	
\resizebox{0.480\textwidth}{!}{\includegraphics{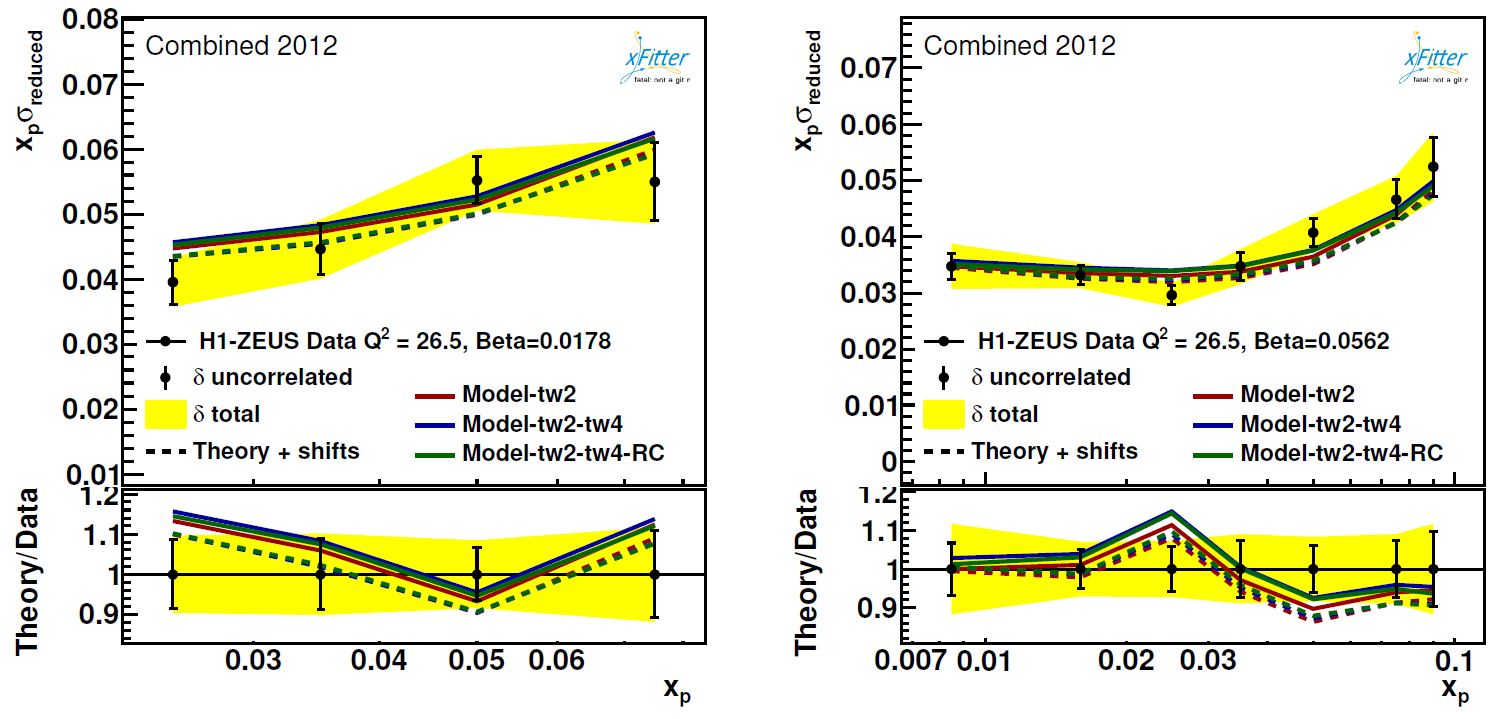}}
\resizebox{0.480\textwidth}{!}{\includegraphics{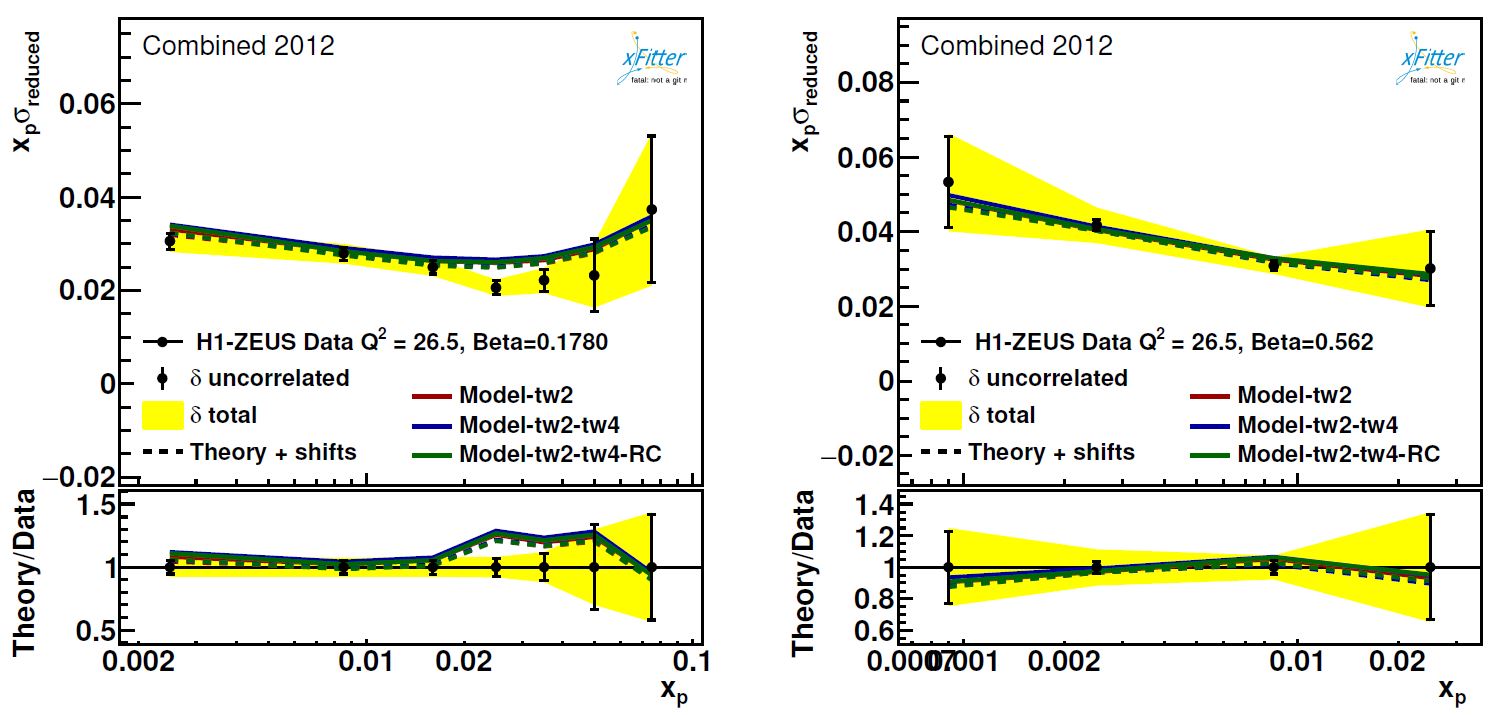}} 	
\resizebox{0.480\textwidth}{!}{\includegraphics{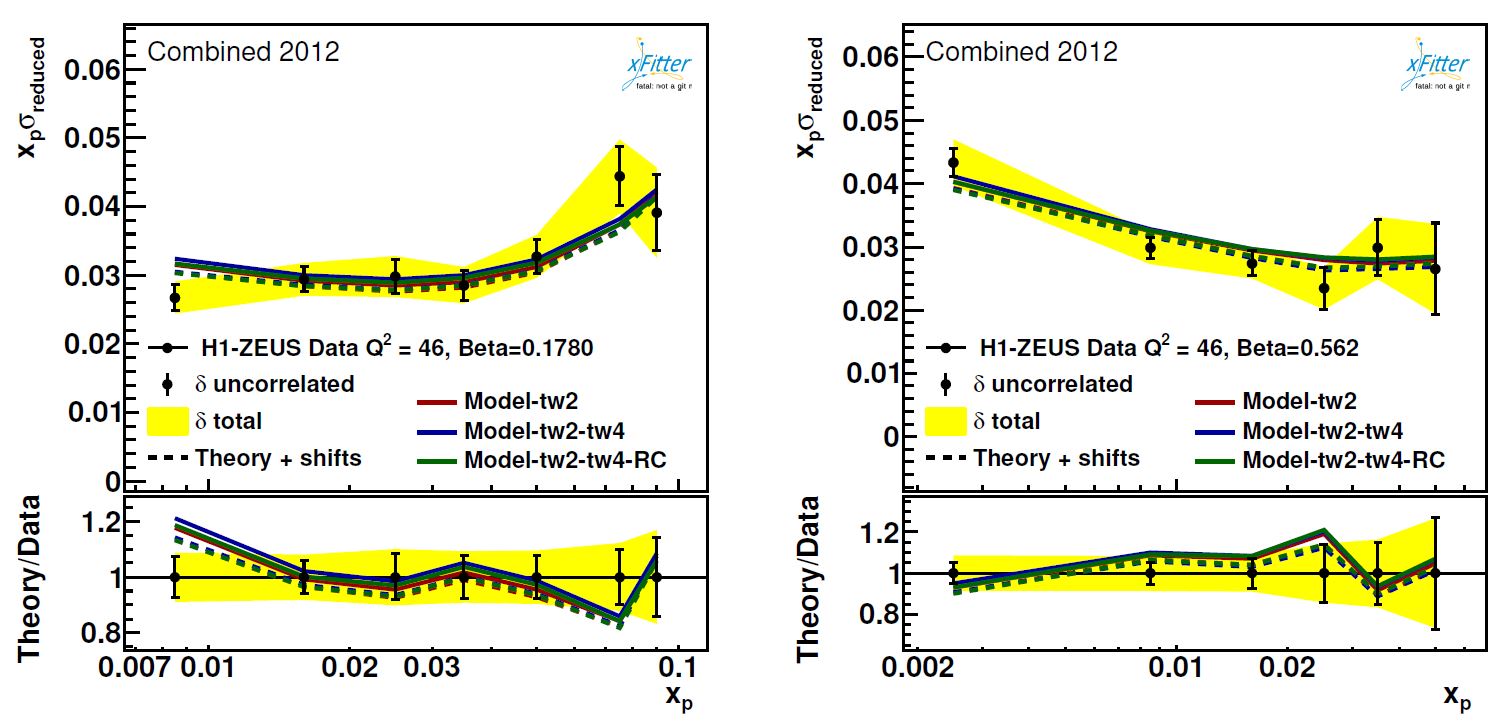}}
\resizebox{0.480\textwidth}{!}{\includegraphics{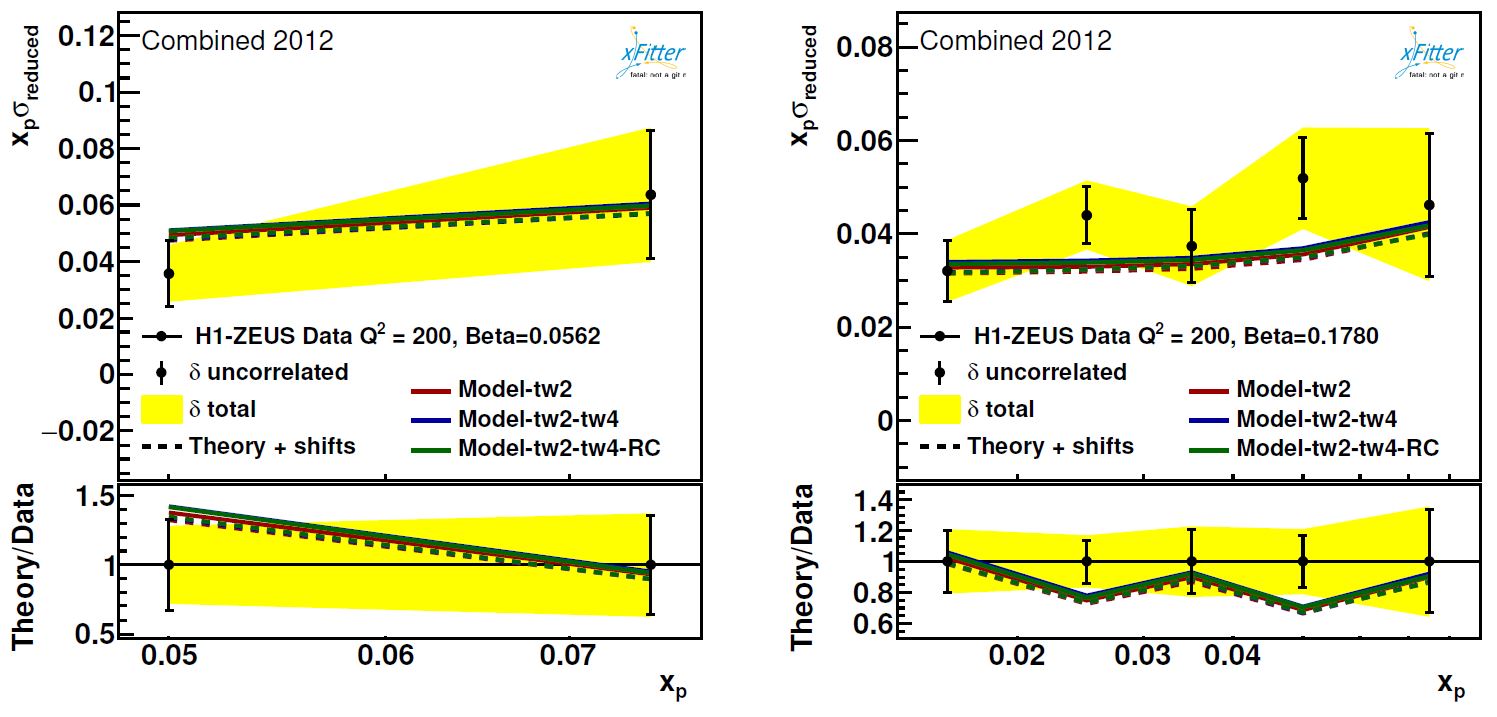}}
\begin{center}
\caption{ \small 
Same as Fig.~\ref{fig:S225} but this time for the  H1/ZEUS combined data.
}
\label{fig:combined}
\end{center}
\end{figure*}

Finally, in Fig.~\ref{fig:combined}, we display a detailed comparison of the 
theory predictions using the three different sets of {\tt SKMHS22} and the 
H1/ZEUS combined data. 
The plots are presented as a function of $x_{\pom}$ and 
for different values of $\beta$ and Q$^2$.  
As one can see, very good agreement is achieved.

As one can see from Figs.~\ref{fig:S225}, 
\ref{fig:H1-LRG-12}  and \ref{fig:combined},  
in general, overall good agreement between 
the diffractive DIS data
and the NLO theoretical predictions are achieved for 
all experiments, which is
consistent with the individual and total $\chi^2$ values 
reported in Table.~\ref{tab:chi2-all}.
Remarkably, the NLO theoretical predictions 
and the data are in reasonable 
agreement in the small- and large-$\beta$ regions, 
and the whole range of $x_{\pom}$.

%
\section{Discussion and Conclusion}\label{Conclusion}

In this work, we performed fits of the diffractive 
PDFs at NLO accuracy in perturbative QCD called {\tt SKMHS22} 
using all available and up-to-date diffractive 
DIS data sets from the H1 and ZEUS Collaborations at HERA. 
We present a mutually consistent set of
diffractive PDFs by adding the high-precision diffractive DIS data
from the H1/ZEUS combined inclusive diffractive cross-sections
measurements to the data sample.

In addition to the standard twist-2 contributions, we also considered 
the twist-4 correction
and Reggeon contribution to the diffractive structure function, 
which dominate in the region of large $\beta$.
The effect of such contributions on the diffractive PDFs and 
cross-sections were carefully examined and discussed. 
The twist-4 correction and Reggeon contribution lead 
to the gluon distribution
which is peaked stronger at high-$\beta$ than in 
the case of the standard twist-2 QCD fit.

The well-established {\tt xFitter} 
fitting methodology, widely used
to determine the unpolarised PDFs, are modified to 
incorporate the higher-twist contributions. 
This fitting methodology is specifically
designed to provide faithful representations of the experimental
uncertainties on the PDFs.

The QCD analysis presented in this work represents the first
step of a broader program.
A number of updates and improvements are foreseen for the future study.

The important limitation of the {\tt SKMHS22}  QCD analysis is
the fact that it is based on the inclusive diffractive DIS cross-section
measurements. 
Despite the fact that the diffractive DIS is the cleanest process for 
the extraction of diffractive PDF, it is scarcely
sensitive to the gluon density.
For the near future, our main aim is to 
include very recent diffractive
dijet production data~\cite{H1:2014pjf,H1:2015okx}, 
which we expect to provide a good
constraint on the determination of the gluon diffractive PDFs.
This will require the numerical implementation
of the corresponding observables at NLO and NNLO accuracy in perturbative QCD 
in the {\tt xFitter} package.
A further improvement for the future {\tt SKMHS22} analyses, 
as a long-term project, is the
inclusion of other observables from hadron 
colliders which could carry some information on flavor separation.

A {\tt FORTRAN} subroutine, which evaluates the three sets 
of {\tt SKMHS22} NLO diffractive PDFs 
presented in this work for
given values of $\beta$, $x_{\pom}$ and Q$^2$ can 
be obtained from the authors upon request. These diffractive PDFs  sets are 
available in the standard {\tt LHAPDF} format.

%
\begin{acknowledgments}
%

Hamzeh Khanpour, Hadi Hashamipour and Maryam Soleymaninia thank the School of Particles and Accelerators, Institute 
for Research in Fundamental Sciences (IPM) for financial support of 
this project. Hamzeh Khanpour also is thankful to the Physics  
Department of University of Udine, and University 
of Science and Technology of Mazandaran for the financial 
support provided for this research.  This work was also supported in part
by the Deutsche Forschungsgemeinschaft (DFG, German Research
Foundation) through the funds provided to the Sino-German Collaborative
Research Center TRR110 ``Symmetries and the Emergence of Structure in QCD''
(DFG Project-ID 196253076 - TRR 110).  The work of UGM was supported in part by the Chinese
Academy of Sciences (CAS) President's International
Fellowship Initiative (PIFI) (Grant No. 2018DM0034) and by VolkswagenStiftung (Grant No. 93562).

\end{acknowledgments}
%

%
%

\clearpage


%

\end{document}